\newif\ifShowKeys
\ShowKeysfalse

\documentclass[11pt]{article}
\pdfoutput=1
\usepackage[usenames,dvipsnames]{xcolor}
\usepackage[setpagesize=false,pagebackref=false, linktocpage, bookmarksopen=true, colorlinks=true, linkcolor=Maroon,citecolor=Maroon,urlcolor=Maroon]{hyperref}

\usepackage{tocloft}

\ifShowKeys \usepackage[notcite]{showkeys} \fi

\usepackage{amsmath, amssymb,amsthm}
\usepackage{amsthm}
\numberwithin{equation}{section}

\usepackage{bm,environ,mathrsfs,array,arydshln}
\usepackage{booktabs,float,slashed,hyperref}
\usepackage[mathcal]{euscript}
\usepackage{tensor} 						
\usepackage{mathabx}
\usepackage{simpler-wick}
\usepackage[vcentermath]{youngtab}

\usepackage{aurical}
\usepackage[T1]{fontenc}
\usepackage[nodayofweek]{date time}
\usepackage{graphicx,epsfig,epic}
\usepackage{tikz} 
\usepackage{tikzfeynman}
\usetikzlibrary{arrows,decorations.pathreplacing,decorations.markings,snakes}
\tikzset{middlearrow/.style={decoration={markings, mark= at position 0.5 with {\arrow{#1}} ,
}, postaction={decorate}}}
\tikzset{decoration={snake,amplitude=.4mm,segment length=2mm,
                       post length=0mm,pre length=0mm}}

\usepackage{framed}						
\definecolor{shadecolor}{rgb}{0.9996078, 0.984314, 0.960784}

\usepackage{comment}

\allowdisplaybreaks

\newcommand{\blue}[1]{\textcolor{blue}{#1}}

\definecolor{myred}{RGB}{233, 33, 45}

\newcommand{\bs}{\begin{shaded}}
\newcommand{\es}{\end{shaded}\noindent}
\def\ba#1\ea{\begin{align}#1\end{align}}		
\newcommand{\be}{\begin{equation}}
\newcommand{\ee}{\end{equation}}
\newcommand{\mc}{\mathcal }

\newcommand{\la}{\label}
\newcommand{\eps}{\varepsilon}

\newcommand{\lp}{\notag \\ & }

\DeclareMathOperator{\tr}{\text{tr}}

\newcommand{\cf}{\textit{cf.} }
\newcommand{\ie}{\textit{i.e.} }
\newcommand{\eg}{\textit{e.g.} }

\newcommand{\N}{\mathcal N}

\makeatletter
\DeclareFontFamily{OMX}{MnSymbolE}{}
\DeclareSymbolFont{MnLargeSymbols}{OMX}{MnSymbolE}{m}{n}
\SetSymbolFont{MnLargeSymbols}{bold}{OMX}{MnSymbolE}{b}{n}
\DeclareFontShape{OMX}{MnSymbolE}{m}{n}{
<-6>  MnSymbolE5
   <6-7>  MnSymbolE6
   <7-8>  MnSymbolE7
   <8-9>  MnSymbolE8
   <9-10> MnSymbolE9
  <10-12> MnSymbolE10
  <12->   MnSymbolE12
}{}
\DeclareFontShape{OMX}{MnSymbolE}{b}{n}{
<-6>  MnSymbolE-Bold5
   <6-7>  MnSymbolE-Bold6
   <7-8>  MnSymbolE-Bold7
   <8-9>  MnSymbolE-Bold8
   <9-10> MnSymbolE-Bold9
  <10-12> MnSymbolE-Bold10
  <12->   MnSymbolE-Bold12
}{}

\let\llangle\@undefined
\let\rrangle\@undefined
\DeclareMathDelimiter{\llangle}{\mathopen}%
 {MnLargeSymbols}{'164}{MnLargeSymbols}{'164}
\DeclareMathDelimiter{\rrangle}{\mathclose}%
 {MnLargeSymbols}{'171}{MnLargeSymbols}{'171}
\makeatother

\catcode`,\active

\catcode`\,12

\def\Xint#1{\mathchoice
   {\XXint\displaystyle\textstyle{#1}}%
   {\XXint\textstyle\scriptstyle{#1}}%
   {\XXint\scriptstyle\scriptscriptstyle{#1}}%
   {\XXint\scriptscriptstyle\scriptscriptstyle{#1}}%
   \!\int}
\def\XXint#1#2#3{{\setbox0=\hbox{$#1{#2#3}{\int}$}
     \vcenter{\hbox{$#2#3$}}\kern-.5\wd0}}

\def\dashint{\Xint-}

\usepackage[object=vectorian]{pgfornament}

\newcommand{\sql}{\sqrt\l}
\renewcommand{\l}{\lambda}
\newcommand{\gym}{g_{\scalebox{0.45}{\text{YM}}}}
\newcommand{\del}{\partial_{\l}}

\newcommand{\vev}[1]{\langle #1 \rangle}
\newcommand{\vevD}[1]{\langle #1 \rangle_{D}}
\newcommand{\VevD}[1]{\bigg\langle #1 \bigg\rangle_{D}}
\newcommand{\NN}{$\mc N=4$ }
\newcommand{\OO}{\mathsf{O}}

\newcommand{\sF}{\mathsf{F}}

\begin{document}


\begin{titlepage}

\vspace*{15mm}
\begin{center}
{\Large\sc   $1/N$ expansion of the D3-D5 defect CFT  }\vskip 9pt
{\Large\sc   at strong coupling }

\vspace*{10mm}

{\large M. Beccaria${}^{\,a}$, A. Cabo-Bizet${}^{\,a}$} 

\vspace*{4mm}
	
${}^a$ Universit\`a del Salento, Dipartimento di Matematica e Fisica \textit{Ennio De Giorgi},\\ 
		and I.N.F.N. - sezione di Lecce, Via Arnesano, I-73100 Lecce, Italy
			\vskip 0.3cm

\vskip 0.2cm {\small E-mail: \texttt{matteo.beccaria@le.infn.it}, \texttt{alejandro.cabo\_bizet@kcl.ac.uk}}
\vspace*{0.8cm}
\end{center}

\begin{abstract}  
\noindent
We consider four dimensional $U(N)$ $\mathcal N=4$ SYM theory interacting with a 3d $\mathcal N=4$ theory living on 
a codimension-one interface and holographically dual to the D3-D5 system without flux. Localization captures several observables in this dCFT, including
its free energy, related to the  defect expectation value, and single trace  $\frac{1}{2}$-BPS composite scalars. These quantities may be computed 
in a hermitian one-matrix model with non-polynomial single-trace potential. We exploit the integrable Volterra hierarchy
underlying the matrix model and systematically study its $1/N$ expansion at any value of the 't Hooft coupling. In particular, the 
strong coupling regime is  determined -- up to non-perturbative exponentially suppressed corrections -- 
by  differential relations that constrain higher order terms in the $1/N$ expansion. The analysis is extended to the model with $SU(N)$ gauge symmetry
by resorting to the more general Toda lattice equations.
\end{abstract}
\vskip 0.5cm
	{
	}
\end{titlepage}

\tableofcontents
\vspace{1cm}

\section{Introduction}

The study of gauge theories with non-trivial low codimensional defects is an active field of research. As usual, 
supersymmetry helps a lot in obtaining exact results. Supersymmetric boundary conditions and interface defects in \NN SYM 
have been studied in \cite{Gaiotto:2008sa,Gaiotto:2008ak}. The $\frac{1}{2}$-BPS defects may be identified with intersecting 
D5- and NS5- branes that share three  directions with a stack of D3-branes where the 4d SYM theory lives.

\medskip
Suitable defects preserving integrability and conformal invariance of \NN SYM allow in principle to 
use the corresponding techniques for integrable models \cite{Beisert:2010jr} and conformal bootstrap \cite{Liendo:2012hy,Liendo:2016ymz,Poland:2018epd}.
Many results have been already obtained by exploiting integrability in dCFT like domain wall versions of \NN SYM, see \eg 
\cite{deLeeuw:2017cop,
deLeeuw:2019usb,
Linardopoulos:2020jck} for  reviews.
The exact formula for tree-level one-point functions in the $SU(2)$ sector have been first 
obtained in \cite{deLeeuw:2015hxa,Buhl-Mortensen:2015gfd}. Extension to the full $SO(6)$ scalar sector has been achieved in 
\cite{deLeeuw:2016umh,deLeeuw:2018mkd} and at one-loop order in \cite{Buhl-Mortensen:2016pxs,Buhl-Mortensen:2016jqo}. 
The closed form determinant formula in the $SU(3)$ sector in \cite{deLeeuw:2016umh} was proven in \cite{DeLeeuw:2019ohp}.

A conjectured all-orders asymptotic (i.e. without wrapping corrections) one-point function formula in the $SU(2)$ sub-sector
has been formulated in  \cite{Buhl-Mortensen:2017ind}. It has been tested in  \cite{Kristjansen:2020mhn}
where an extension to gluonic and fermionic sectors (in presence of fluxes) is also treated. Further results on one-point functions from the point of view of the 
overlap between standard Bethe eigenstates and boundary states of the integrable super spin chains may be found in \cite{Kristjansen:2020vbe,Kristjansen:2021xno}.
Recently, classical integrability of the D3-D5 brane system was proven on the string theory side in \cite{Linardopoulos:2021rfq}, complementing
the earlier work \cite{Dekel:2011ja}.

\medskip
The bootstrap approach for correlation functions of local operators may be formulated in the presence of boundaries or domain 
walls that preserve the conformal sub-algebra longitudinal to the defect \cite{Liendo:2012hy,Billo:2016cpy,Liendo:2016ymz}. 
The structure is richer than in the standard setup since additional structure constants appear, intrinsic to the defects, and defining the so-called 
defect one-point functions of bulk local operators. They are the necessary data, with the usual OPE expansion of  local operators, to 
determine (local) correlation functions in the presence of the conformal defect. For recent results, see \cite{Gombor:2020auk}.

\medskip
At codimension one, we deal with boundaries or interface defects, the two being typically linked by a suitable (un)folding construction
relating half-space to full space bisected by a domain wall defect. In this paper, we will consider the so-called D3-D5 system
\cite{Constable:1999ac,Karch:2000gx,DeWolfe:2001pq,Erdmenger:2002ex}.
Its IIB string side description involves D3- and D5-branes sharing the 012 directions.
The $N$ D3-branes fill the 1234 directions, while the D5-brane spans 234890. All the branes are placed at the origin of the transverse coordinates.
In the near horizon limit, this system has 
geometry $\text{AdS}_{5}\times \text{S}^{5}$ split in two  by the probe D5-brane with world-volume geometry $\text{AdS}_{4}\times \text{S}^{2}$. 
The near horizon geometry of the D3-D5 system contains the 
closed IIB superstrings excitations in $\text{AdS}_{5}$, dual to \NN SYM states on the $\mathbb R^{4}$ boundary of AdS$_{5}$.
Also, we have open strings connecting the D3- and D5-branes,  dual to field theory excitations on the $\mathbb R^{3}$ boundary of AdS$_{4}$, to be regarded
as a codimension one defect in $\mathbb R^{4}$. 

\medskip
The dual field theory  is \NN SYM in $3+1$ dimensions and interacting with a 3d \NN field theory living on the  $2+1$-dimensional defect. 
The system of  bulk SYM plus the interface at $x_{1}=0$ has a reduced symmetry since translation  in the $x_{1}$ direction is broken, 
together with rotations and boosts in the Lorentz $SO(1,3)$ involving that direction. Thus $SO(1,3)\to SO(1,2)\simeq SU(2)$. Conformal transformations are those preserving $x_{1}=0$ (dilatations and three special conformal
transformations) so $SO(4,2)\to SO(3,2)$. Supersymmetry is also partially broken since supercharges anticommute into translations. One finds that 
the defect breaks half of the supercharges and superconformal charges, i.e. preserves a $\frac{1}{2}$-BPS subalgebra of the full superconformal
algebra.  In more details, the R-symmetry algebra is reduced $\mathfrak{su}(4)\to  \mathfrak{so}(3)\oplus\mathfrak{so}(3)$ and 
the superconformal algebra $\mathfrak{psu}(2,2|4)\to \mathfrak{osp}(4|4)$. \footnote{
A conformal defect of codimension $p$ in 4f CFT breaks the conformal group to subgroups of $SO(4,2)\to SO(4-p,2)\times SO(p)$. 
In supersymmetric theories the superconformal algebra is reduced to BPS subalgebras, \ie preserving a certain number of supercharges,  and the maximally supersymmetric ones are $\frac{1}{2}$-BPS.
For $\mathfrak{psu}(2,2|4)$, they are classified in \cite{DHoker:2008wvd}. The codimension-one case (interface) has reduced supersymmetry $\mathfrak{osp}(4|4, \mathbb R)$
and the interface has 3d $\mathcal N=4$ supersymmetry on its world-volume. This super-algebra contains $\mathfrak{so}(3)\oplus \mathfrak{so}(3)\oplus \mathfrak{so}(3,2)$, 
\ie a maximal subalgebra of the R-symmetry $\mathfrak{so}(6)$ and the conformal algebra on the interface.}

\medskip
This basic setup may be generalized when  the worldvolume gauge fields
of the probe D5-brane have a monopole bundle with quantized $U(1)$ magnetic flux.
Now, in the field theory dual, the defect separates $3+1$-dimensional space-time into regions where the  gauge group of \NN SYM has different ranks. This corresponds to the situation where $k$ of the 
$N$ D3-branes end on the worldvolume of a D5-brane
and $k$ turns out to be the number of units of Dirac monopole flux, see Fig.~\ref{fig:string-setup}.
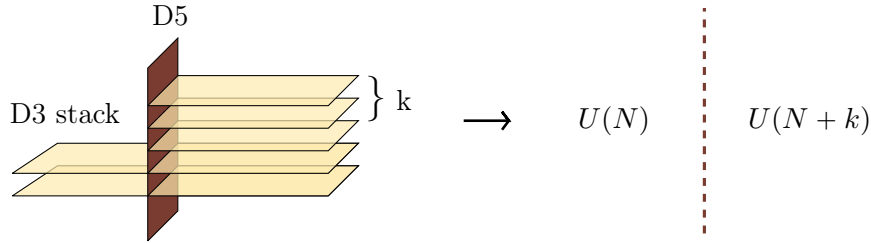
\begin{figure}[t]
%
\definecolor{d3brane}{RGB}{254,235,176}
\definecolor{d5brane}{RGB}{122,60,49}
\definecolor{arrow}{RGB}{146,179,233}

\begin{center}
\begin{tikzpicture}[line width=1 pt]
\begin{scope}[shift={(-1,0)}]
	\begin{scope}[shift={(0,0)}]     \draw [thin, black, fill=d3brane, fill opacity = 0.7] (-2.2,-2) -- (-2+0.4,-2+0.4) -- (2+0.4,-2+0.4) -- (2,-2) -- (-2.2,-2) ;  \end{scope}
	\begin{scope}[shift={(0,0.3)}]  \draw [thin, black, fill=d3brane, fill opacity = 0.7] (-2.2,-2) -- (-2+0.4,-2+0.4) -- (2+0.4,-2+0.4) -- (2,-2) -- (-2.2,-2) ;   \end{scope}

	\draw [thin, black, fill=d5brane] (-0.4,-2.6) -- (-0.4,-0.3) -- (-0.4+0.4,-0.3+0.4) -- (-0.4+0.4,-2.6+0.4) --(-0.4,-2.6) ;

	\begin{scope}[shift={(0,0)}]    \draw [thin, black, fill=d3brane, fill opacity = 0.7] (-0.4,-2) -- (-0.4+0.4,-2+0.4) -- (2+0.4,-2+0.4) -- (2,-2) -- (-0.4,-2) ;   \end{scope}
	\begin{scope}[shift={(0,0.3)}]  \draw [thin, black, fill=d3brane, fill opacity = 0.7] (-0.4,-2) -- (-0.4+0.4,-2+0.4) -- (2+0.4,-2+0.4) -- (2,-2) -- (-0.4,-2) ; \end{scope}
	\begin{scope}[shift={(0,0.6)}] \draw [thin, black, fill=d3brane, fill opacity = 0.7] (-0.4,-2) -- (-0.4+0.4,-2+0.4) -- (2+0.4,-2+0.4) -- (2,-2) -- (-0.4,-2)  ;\end{scope}
	\begin{scope}[shift={(0,0.9)}]\draw [thin, black, fill=d3brane, fill opacity = 0.7] (-0.4,-2) -- (-0.4+0.4,-2+0.4) -- (2+0.4,-2+0.4) -- (2,-2) -- (-0.4,-2)  ; \end{scope}
	\begin{scope}[shift={(0,1.2)}]\draw [thin, black, fill=d3brane, fill opacity = 0.7] (-0.4,-2) -- (-0.4+0.4,-2+0.4) -- (2+0.4,-2+0.4) -- (2,-2) -- (-0.4,-2)  ; \end{scope}
	
	\node at (-1.5,-0.9) {D3 stack};
	\node at (-0.1,0.4) {D5};
	\node at (2.8,-0.7) {{\LARGE \}} k};
\end{scope}

\begin{scope}[shift={(2.8,-1)},scale=0.3]
	\draw (0.0,0.0) -- (2.0,0.0) -- (1.7,0.4) -- (2.0,0.0) -- (1.7,-0.4);
\end{scope}

\begin{scope}[shift={(6,0)}]
	\draw [line width=1.2, dashed, d5brane] (0.0,-2.5) -- (0.0,0.5);
	\node at (1.4,-1) {$U(N+k)$};
	\node at (-1.2,-1) {$U(N)$};
\end{scope}
\end{tikzpicture}
\end{center}
\caption{String setup for the D3-D5 system with flux, and corresponding dual dSYM theory.}
\la{fig:string-setup}
\end{figure}
At large $N$, this is same as 
considering a single D5-brane being the end locus for $N + k$  D3-branes for $x_{1}\to 0^{+}$ and $N$ D3-branes for $x_{1}\to 0^{-}$. 

\medskip
While the D5-brane interface for $k=0$ has a simple Lagrangian description, as a transparent interface stacked with a 
bifundamental hypermultiplet on its worldvolume, this is not the case when $k > 1$ 
which involve the singular Nahm pole boundary condition \cite{Gaiotto:2008sa}. The case $k=1$ is somewhat in between and
smoothness across it has been recently fully clarified in \cite{Kristjansen:2020mhn}. \footnote{In the field theory, the interface
may be realized by  assigning a non zero expectation value to three \NN SYM scalars in the half space $x_{1} > 0$, in the Nahm pole form 
$\vev{\Phi_{i=1,2,3}}  = t_{i}/x_{1}$ where $t_{i}$ is an irreducible $k$-dimensional
representation of $\mathfrak{su}(2)$  \cite{Nahm:1979yw,Diaconescu:1996rk,Karch:2000gx}.  This approach trivializes
for $k=1$ and require ad hoc boundary conditions.}
Finally, we remark that the opposite limit 
$k\gg 1$ may facilitate the comparison with the string side, see \eg  \cite{Nagasaki:2012re}.
Integrability of the D3-D5 system holds in the usual sense of describing the spectrum of anomalous dimensions in terms of 
an integrable spin chain Hamiltonian and has been discussed in \cite{DeWolfe:2004zt} for $k=0$ and generalized to any $k$
in \cite{Ipsen:2019jne}.

\medskip
In this paper, we consider the localization matrix model that captures the $k=0$ D3-D5 system \cite{Robinson:2017sup,Wang:2020seq,Komatsu:2020sup}
and discuss the all-orders $1/N$ expansion of specific observables, \ie the defect interface expectation value playing the role of free energy for this system, 
and the one-point functions of certain BPS scalar primaries in the presence of the defect. The key observation is that the defect modifies the 
\NN SYM Gaussian matrix model by introducing a single-trace non-polynomial potential. The modified matrix model is related to the Toda integrable
hierarchy and its Volterra reduction. This allows to derive systematically  its $1/N$ expansion in terms of exact functions of the 't Hooft coupling. 
\footnote{As a remark, the analysis of $1/N$ corrections 
based on the underlying integrable hierarchies turns out to be much more effective than the general purpose topological recursion based on loop equations, see for instance
 \cite{Beccaria:2020ykg,Beccaria:2021alk}.}

\medskip
This strategy has been recently applied in \cite{Beccaria:2021ism,Beccaria:2022kxy}
to the four-dimensional $\mc N=2$ superconformal $Sp(2N)$ gauge theory containing the vector multiplet coupled 
to four hypermultiplets in  the fundamental representation and one hypermultiplet in the rank-2 antisymmetric representation. The special feature of this model 
is that its localization matrix model has an interaction potential containing single-trace terms only. Application of the Toda lattice equations solves the $1/N$ expansion 
at any 't Hooft coupling, including the strongly coupled regime where the gauge theory can be compared with its string dual, an orientifold of $AdS_{5}\times S^{5}$ type IIB string.
A similar analysis is possible here precisely because of the absence of double (or higher ) trace interactions. Such terms are typical in $\mc N=2$ models, 
see for instance \cite{Beccaria:2020hgy,Beccaria:2021hvt,Beccaria:2021vuc,Beccaria:2021ksw}, and greatly complicate the analysis, altough exact strong coupling 
results are available in some cases  \cite{Beccaria:2022ypy}.  

\medskip
The plan of the paper is the following. In Section \ref{sec:loc}, we briefly recall the localization approach to the fluxless D3-D5 system, the resulting single-trace 
matrix model, and the observables we consider. These are the free energy, \ie logarithm of partition function with defect insertion, and certain $\frac{1}{2}$-BPS local 
scalar operators in the localization relevant cohomology. In Section \ref{sec:free}, we present some preliminary results on the $1/N$ expansion of the free energy for the model
with gauge group $U(N)$. A direct analysis shows the existence of peculiar differential relations expressing the higher order $1/N$ corrections $\sF_{n}(\l)$ to the free energy ($\l$ being the
't Hooft coupling)
to the leading term $\sF_{0}(\l)$. In Section \ref{sec:hier}, we recall the connection between single-trace hermitian one-matrix models and integrable
hierarchies, in particular the Toda lattice and its Volterra reduction. In \ref{sec:free-exp}, we apply these structures to generate systematically the 
differential relations giving all $1/N$ corrections to the free energy. We also analyse the strong coupling  expansion of the functions $\sF_{n}(\l)$
deriving the resummation of the terms with highest power of $\l$ at each order in the genus expansion. In Section \ref{sec:free-sun} and \ref{sec:sun}, we extend our analysis
to the model with gauge group $SU(N)$. Technically, this is non-trivial since imposing a traceless condition is not natural from the point of view of the 
integrable hierarchies. This is the counterpart of the physical fact that the extra $U(1)$ degrees of freedom in $U(N)$ case do not decouple from the defect.
Nevertheless, we provide exact formulas for the $1/N$ corrections up to the order $1/N^{3}$. 
In the remaining part of the paper we move on the one-point function of the BPS scalars. In Section \ref{sec:one-un}, we provide some preliminary 
perturbative results for the operators with lower dimension, working out the corrections up to (relative) $1/N^{2}$. Note that the $1/N$ expansion 
is not only in even powers of $1/N$ unlike what happens in simple defectless \NN SYM. Section \ref{sec:one-volterra} is devoted to an analytic derivation of the 
exact $1/N$ expansion, \ie valid at all coupling, by exploiting the Volterra hierarchy in the $U(N)$ model. Again, all the computed corrections are 
studied at strong coupling by exploiting the exact differential relations. The extension to the model with $SU(N)$ symmetry is presented in Section \ref{sec:one-sun}.
Conclusions and open issues are summarized in Section \ref{sec:conclusions}. Finally, several technical appendix sections are included providing further details and proofs.

\section{Localization and defect CFT}
\la{sec:loc}

In presence of enough supersymmetry and in definite sectors of observables, we may use localization \cite{Pestun:2016zxk}
to compute one-point functions of protected operators,  which are definitely non-trivial objects in dCFT  \cite{Robinson:2017sup, Wang:2020seq, Komatsu:2020sup}.
As first conjectured in \cite{Drukker:2007qr,Drukker:2007yx}, and later proved in \cite{Pestun:2009nn} by localization, 
it is possible to show that \NN SYM  $\frac{1}{2}$-BPS Wilson loops  restricted to a two-sphere ${\rm S}^{2}_{\rm YM}$ are described by a bosonic
2d YM theory. In the localization proof, one chooses a particular supercharge $\mc Q$ of the 4d SYM which is nilpotent when restricted to ${\rm S}^{2}_{\rm YM}$
and shows that the 2d YM emerges as an effective description of the $\mc Q$-cohomology in the original SYM. The precise 2d/4d dictionary has been 
clarified in \cite{Giombi:2009ds} and many applications followed, as the check of AdS/CFT described in  
\cite{Giombi:2009ms, Bassetto:2009rt, Gerchkovitz:2016gxx, Giombi:2018qox,Giombi:2018hsx}.

\medskip
The classification of general conformal defects of the 4d \NN SYM in the $\mc Q$-cohomology
has been accomplished in \cite{Wang:2020seq}, including in particular domain walls or boundaries. 
A BPS interface crosses the previous ${\rm S}^{2}_{\rm YM}$ at an equator ${\rm S}^{1}$ that is a codimension-one defect in the 2d YM.
The $\mc Q$-cohomology is extended to take into account local insertions on ${\rm S}^{1}$. The 2d YM turns out to be 
non-trivially coupled to a certain one-dimensional topological quantum
mechanics on ${\rm S}^{1}$ 
\cite{Chester:2014mea, Beem:2016cbd, Dedushenko:2016jxl,Dedushenko:2017avn,Dedushenko:2018icp} 
defining to so-called defect YM (dYM).
A large class of defect observables in
the SYM that preserve a common supercharge $\mc  Q$ have simple descriptions in the dYM sector and their correlation functions can be extracted using 2d 
gauge theory techniques, see \cite{Blau:1993hj, Cordes:1994fc} for reviews.
 
\medskip
For the D5-brane interface that interpolates between $U(N)$ and 
$U(N + k)$ SYM theories for $k\ge 0$, the dYM sector has been determined in \cite{Wang:2020seq,Komatsu:2020sup}
for any $k$ by using S-duality of the bulk SYM theory and the related mirror
symmetry acting on the boundary conditions. 
By two-dimensional gauge theory techniques in the dYM effective theory, the computation
of the defect one-point function $\vevD{ \mc O}$ is reduced to a single-matrix integral. Compared to the
simple Gaussian matrix model familiar for SYM, the relevant matrix model involves a novel single trace
potential, which comes from the D5-brane defect. By solving this matrix model
in the planar large N limit, it is possible to  determine the one-point functions $\vevD{\mc O}$ as exact functions 
of the 't Hooft coupling $\l = g^{2}N$.

\subsection{The matrix model for the $k=0$ $U(N)$ D3-D5 system}

Let us sketch the construction of the matrix model we need for the $k=0$ D3-D5 system. Let us begin by 
briefly recalling what happens in the  case without defects. Let us consider the following 2-sphere ${\rm S}^{2}_{\rm YM}$ (we implicitly stereographically map 
$\mathbb R^{4}$ to the sphere $S^{4}$ of radius $R$)
\be
x_{4}=0, \qquad x_{1}^{2}+x_{2}^{2}+x_{3}^{2} = R^{2}.
\ee
Localization can be performed with respect to a particular supercharge $\mc Q$ such that the BPS locus $\mc Q \Psi=0$ is parametrized 
by a twisted connection $\mc A$ on $\rm S^{2}_{\rm YM}$ (depending on both $A$ and three scalar fields of \NN SYM) and the SYM action on ${\rm S}^{4}$ reduces 
on the BPS locus to 2d constrained (\ie at zero instanton number) 
YM action on ${\rm S}^{2}_{\rm YM}$ with imaginary coupling $g^{2}_{2} = -\gym^{2}/(2\pi\,R^{2})$. \footnote{
\la{foot:sym10}
The 10d SYM action is
$S = -\frac{1}{2\gym^{2}}\int d^{4}x\, \tr(\frac{1}{2}F_{MN}F^{MN}-\Psi\Gamma^{M}D_{M}\Psi)$. The 10d spacetime indices split into 4d indices
$\mu=1,2,3,4$ and R-symmetry indices $a=5,\dots, 9, 0$. The gauge field $A_{M}$ contains the 4d gauge field $A_{\mu}$ and six scalars $\Phi_{a}$.
The gaugino is a chiral spinor of Spin(10). Finally $D = d+A$.}
Observables are in the $\mc Q$ cohomology and include $\frac{1}{8}$-BPS Wilson loops and certain local operators on ${\rm S}^{2}_{\rm YM}$, see \cite{Giombi:2009ds}.

\medskip 
Let us now consider \NN SYM with a codimension one interface at $x_{1}=0$. As we mentioned, it is realized by a single D5-brane along the 234890 directions  
intersecting $N$  D3-branes along the 1234 directions in the 10d  spacetime. 
On the $\mathfrak{osp}(4|4)$ preserving interface, the \NN vector multiplet splits into 3d \NN multiplets, the hypermultiplet $(A_{1}, X_{a}, \Psi_{-})$
and the vector multiplet $(A_{2,3,4}, Y_{a}, \Psi_{+})$, where $\Psi_{\pm}$ are suitable Majorana projections of $\Psi$ and 
$\mathsf{X} = (\Phi_{8}, \Phi_{9}, \Phi_{0})$, $\mathsf{Y} = (\Phi_{5}, \Phi_{6}, \Phi_{7})$ in the notation of footnote \ref{foot:sym10}.
In the half space $\mathbb R^{4}_{+}$ one can consider two 
types of supersymmetry preserving boundary conditions (and their possible mixing), \ie D5-brane type (or generalized Dirichlet)  or 
NS5-brane type that are related to the previous ones by S-duality \cite{Gaiotto:2008ak}. They read (we show only
bosonic fields)
\ba
\text{D5:} &\qquad \left. D_{1}X_{i}-\tfrac{1}{2}\eps_{ijk}[X_{j}, X_{k}]\right|_{x_{1}=0} = 
\left. Y_{i} \right|_{x_{1}=0}=0, \notag \\
\text{NS5:} & \qquad \left. X_{i}\right|_{x_{1}=0} = \left. D_{1}Y_{i}\right|_{x_{1}=0} = 0.
\ea

\medskip 
At $x_{1}=0$, the interface has 3d \NN superconformal symmetry on its 234 worldvolume. It contains a non-trivial
1d sector described by topological quantum mechanics living on the circle $x_{1} = x_{4} = 0$, $x_{2}^{2}+x_{3}^{2} = R^{2}$.
This is the boundary of $HS^{2}_{\rm YM}$ and appears in the $\mc Q$ cohomology in presence of the defect.

\medskip 
When we give D5-type boundary conditions
to the hypermultiplet and NS5-type to the vector multiplet, then the vector multiplet can be coupled to 3d \NN matter SCFT, see for instance 
\cite{DeWolfe:2004zt} for the explicit Lagrangian in component fields.
At the level of the 2d (constrained) YM theory, this gauges the topological quantum mechanics on the intersection between ${\rm S}^{2}_{\rm YM}$
and the interface $x_{1}=0$.

\medskip 
The detailed analysis in  \cite{Wang:2020seq} confirms (and  clarifies in full generality) the one-matrix model
first proposed in \cite{Robinson:2017sup}. This may be quickly obtained, with some hindsight,
by a combination of  the results for 4d \NN SYM
and independent localization computations in (intrinsically) three-dimensional Chern-Simons-matter theories \cite{Kapustin:2009kz}.
The partition function is indeed as in \NN SYM, but with a rather simple defect term associated with the 3d fundamental 
hypermultiplet living at the defect. It reads \cite{Robinson:2017sup,Wang:2020seq,Komatsu:2020sup}
\ba
\la{2.3}
Z_{N}(\gym) = 
\int_{\mathfrak{u}(N)} \mc DM\,\,\exp\bigg[-\frac{8\pi^{2}}{\gym^{2}}\,\tr M^{2}-\tr\log [2\,\cosh (\pi M)]\bigg],
\ea
where details on the measure, \ie normalization, will be discussed in a moment.

\medskip 
According to the 4d/2d dictionary, and up to a normalization, the matrix $M$ is associated with the (scalar) $\frac{1}{8}$-BPS chiral primary
(see also \cite{Drukker:2009sf})
\be
\la{2.4}
\tr M\leftrightarrow \tr \Phi = \tr(x_{1}\,\Phi_{7}+x_{2}\,\Phi_{9}+x_{3}\,\Phi_{0}+i\,\Phi_{8}).
\ee
as proved by Giombi and Pestun in  \cite{Giombi:2009ds} \footnote{
The field in the r.h.s. of (\ref{2.4}) is $\mc Q$-closed and any gauge invariant 
functional of it may be treated in a similar way. Also, the general discussion in \cite{Wang:2020seq} gives the theoretical framework to deal with the 
full set of states in the $\mc Q$-cohomology, including disorder operators.
}
If the defect is placed at, say,  $x_{1}=0$ in the $(x_{1},x_{2},x_{3},x_{4})$ half-space 
$\mathbb R_{+}^{4}$, a generic scalar operator with definite conformal weight has one-point function (in some conventional  normalization)
\be
\la{2.5}
\vev{\mc O(x)}_{\mathbb R^{4}_{+}} = \frac{h_{\mc O}}{|x_{1}|^{\Delta_{\mc O}}}.
\ee
We are interested in the specialization where $\mc O$ is a protected $\frac{1}{2}$-BPS operator placed at $(1,0,0,0)$
and belonging to the cohomology of the charge $\mc Q$ used for localization. In this case, the one-point function (\ref{2.5}) reduces to 
$\vev{\mc O(x)} = h_{\mc O}$.
A well known case, discussed in  \cite{Wang:2020seq,Komatsu:2020sup} is that of 
single trace composite scalars of the form 
\be
\la{2.6}
\mc O_{p}=\tr \Phi^{p},
\ee
that transform in the $[0,p,0]$ representation of the R-symmetry algebra $\mathfrak{su}(4)$ and whose 
one-point function in (\ref{2.5}) is non-vanishing only when $p$ is even. 
\footnote{
For related investigations of different correlators in the same BPS sector, see also \cite{Dedushenko:2020vgd,Dedushenko:2020yzd}.
}
For these composite operators,  as in the non-defect case, one has also to relate the sphere computation
to flat space by disentangling the operator mixing due to the fact that  the regulated theory on $S^{4}$ breaks the $U(1)_{R}$ symmetry 
and mixing among operators with different  $R$-charge is possible \cite{Gerchkovitz:2016gxx,Gomis:2015yaa}.
This amounts to a normal ordering prescription $\mc O\to :\mc O:$, see for instance \cite{Billo:2017glv,Rodriguez-Gomez:2016cem}.
The associated one-point functions are computed by inserting 
$:\tr M^{p}:$ in the one-matrix model (\ref{2.3}), where mixing has to be computed
using the \NN $U(N)$ SYM matrix model without the defect contribution.

Before discussing the relevant observables considered in this paper, let us give (\ref{2.3}) in full detail. 
In the general case with flux, we have gauge groups $U(N)$ and $U(N+k)$ on the two sides of the interface. The defect partition function is, \cf Eq.~(2.33) of \cite{Komatsu:2020sup},
\be
\la{2.7}
Z_{N,k}^{\rm D5}(\gym) = \pi^{N}\left(\frac{4\pi}{\gym^{2}}\right)^{\frac{N^{2}+(N+k)^{2}}{2}}\,e^{\frac{4\pi}{\gym^{2}}\frac{k(k^{2}-1)}{24}}G(1+k)\,Z_{N,k}(\gym),
\ee
with 
\ba
\la{2.8}
Z_{N,k}(\gym) &= \frac{(-i)^{Nk}}{N!}\int \prod_{n=1}^{N}\frac{dx_{n}}{2\pi}\frac{\prod_{1\le n<m\le N}(x_{n}-x_{m})^{2}}{\prod_{n=1}^{N}\cosh \pi(x_{n}+\frac{ik}{2})}
\prod_{n=1}^{N}\prod_{s=-\frac{k-1}{2}}^{\frac{k-1}{2}}(x_{n}-is)\,e^{-\frac{8\pi^{2}}{\gym^{2}}\sum_{n=1}^{N}x_{n}^{2}}.
\ea
Up to a non-trivial $\gym$-dependent normalization, this takes the form (\ref{2.3}) for $k=0$. This (simplest) case is 
still non-trivial. We will write \footnote{See Appendix \ref{app:norm} for comments on the prefactor in (\ref{2.9}).}
\be
\la{2.9}
Z_{N,0}^{\rm D5} =  \pi^{N}\left(\frac{4\pi}{\gym^{2}}\right)^{N^{2}}\, Z_{N},
\ee
where $Z_{N}$ is a specialization of the general single-trace partition function 
\ba
\la{2.10}
Z_{N}(\bm t) &= \int_{\mathfrak{u}(N)}\mc DM\, e^{-\tr W(M, \bm t)} = \frac{1}{N!}\int\prod_{n=1}^{N}\frac{dx_{n}}{2\pi}\,\prod_{1\le n<m\le N}(x_{n}-x_{m})^{2}\,e^{-\sum_{n=1}^{N}W(x_{n}; \bm t)},
\ea
with a generic multi-coupling potential 
\ba
W(x; \bm{t}) &= \sum_{n=1}^{\infty}t_{n}\, x^{n}.
\ea
Comparing (\ref{2.10}) with (\ref{2.8}), the ($k=0$) D3-D5 system corresponds to the total single-trace potential
\ba
\la{2.12}
W(x) &= \frac{8\pi^{2}}{\gym^{2}}x^{2}+L(x), 
\ea
with the specific form of the function $L(x)$
\ba
\la{2.13}
L(x) &= \log\cosh(\pi x) = \sum_{n=1}^{\infty}\frac{2^{2n}(2^{2n}-1)B_{2n}\pi^{2n}}{2n(2n)!}x^{2n},
\ea
where $B_{2n}$ are Bernoulli numbers.
Notice that the quadratic term in the small $x$ expansion of $L(x)$ combines with $\frac{8\pi^{2}}{\gym^{2}}$ to give the full $t_{2}$ coupling, \ie
\be
t_{2n} =  \frac{8\pi^{2}}{\gym^{2}}\,\delta_{n,2}+\frac{2^{2n}(2^{2n}-1)B_{2n}\pi^{2n}}{2n(2n)!}.
\ee

\subsection{Free energy and one-point functions}

The free energy of the matrix model (\ref{2.10}) is  the  expectation value of the defect insertion $D = \exp[-\tr L(M)]$ in the SYM matrix model.
Thus,  
\be
F_{N}(\gym) = -\log \vev{D}_{\rm SYM},
\ee
where we choose the normalization in order to have $F=0$ without the defect, \ie
\be
\vev{f(M)}_{\rm SYM} = \int DM f(M) e^{-\frac{8\pi^{2}}{\gym^{2}}\tr M^{2}}/\int DM  e^{-\frac{8\pi^{2}}{\gym^{2}}\tr M^{2}},
\ee
so that $\vev{1}_{\rm SYM}=1$. Notice that the prefactor in (\ref{2.9}) should be included,
but for the purposes of the next sections it will be convenient to adopt the above simpler normalization.
In the 't Hooft limit, with fixed $\l=\gym^{2}N$ and large $N$, the free energy admits the $1/N$ expansion \footnote{We adopt a little abuse of language and 
use the same symbol to denote the free energy when expressed in terms of $\l$ in the 't Hooft limit.}
\be
\la{2.17}
\Delta F_{N}(\l) = N\,\sF_{0}(\l)+\sF_{1}(\l)+\frac{1}{N}\sF_{2}(\l)+\cdots,
\ee
Defect one-point functions are defined as \footnote{Of course the matrix model 
normalization drops in (\ref{2.18}).}
\be
\la{2.18}
\vevD{\mc O} \equiv \frac{\vev{\mc O\,\, D}_{\rm SYM}}{\vev{D}_{\rm SYM}}.
\ee
We will focus on the case $\mc O = :\mc O_{n}: = :\tr M^{n}:$ and denote its   
large $N$ expansion at fixed $\l = \gym^{2}N$ by
\ba
\la{2.19}
\OO_{n}(\l; N) &=  \vevD{ : \tr M^{n} : }  = \OO_{n}^{(0)}(\l)+\OO_{n}^{(1)}(\l)\frac{1}{N}+\OO_{n}^{(2)}(\l)\frac{1}{N^{2}}+\cdots\, .
\ea
For weak coupling calculations it will be often convenient to rescale the matrix $M$ according to 
\be
\la{2.20}
A = \sqrt\frac{8\pi^{2}}{\gym^{2}}\,M.
\ee
The  associated multi-trace operators defined by 
\be
\Omega_{\bm{n}}  = \tr A^{n_{1}}\,\tr A^{n_{2}}\cdots \tr A^{n_{K}}, \qquad |\bm n| = n_{1}+\cdots + n_{K},
\ee
obey 
\be
\la{2.22}
\mc O_{\bm{n}}(M) = \left(\frac{\l}{8\pi^{2}N}\right)^{|\bm n|/2}\,\Omega_{\bm{n}}(a),\quad \text{and}\quad :\mc O_{\bm{n}}(M): = \left(\frac{\l}{8\pi^{2}N}\right)^{|\bm n|/2}\,:\Omega_{\bm{n}}(a):~.
\ee
In terms of the new matrix $A$, the partition function (\ref{2.10}) reads
\ba
\la{2.23}
Z_N(\bm t) &= \left(\frac{\gym^{2}}{8\pi^{2}}\right)^{N^{2}/2}\, \int_{\mathfrak{u}(N)} \mc DA\, \exp\bigg[-\tr A^{2}-\tr L\left(\sqrt{\frac{\gym^{2}}{8\pi^{2}}}\,A\right)\bigg].
\ea

\section{Free energy $1/N$ expansion in the $U(N)$ model: direct method}
\la{sec:free}

Evaluating perturbatively  (\ref{2.23}), we can obtain  the free energy $F_{N}(\l)$ at weak coupling and with the specific D3-D5 function $L$ in (\ref{2.12}).
The expansion of the first coefficient functions $\sF_{n}(\l)$ in (\ref{2.17}) are
\ba
\la{3.1}
\sF_{0}(\l) &= \frac{\l }{32}-\frac{\l ^2}{1536}+\frac{\l 
^3}{36864}-\frac{17 \l ^4}{11796480}+\frac{31 \l 
^5}{353894400}-\frac{691 \l ^6}{118908518400}+\cdots, \notag \\
\sF_{1}(\l) &= -\frac{\l ^2}{1024}+\frac{\l ^3}{12288}-\frac{11 \l 
^4}{1572864}+\frac{29 \l ^5}{47185920}-\frac{83 \l 
^6}{1509949440}+\cdots, \notag \\
\sF_{2}(\l) &= -\frac{\l ^2}{3072}+\frac{7 
\l ^3}{73728}-\frac{35 \l ^4}{2359296}+\frac{563 \l 
^5}{283115520}-\frac{1681 \l ^6}{6794772480}+\cdots\, .
\ea
The same computation can be done while keeping generic couplings $\bm t$ in the even potential $L$, \footnote{So, we \underline{do not} include in $t_{2}$ the free action contribution $\frac{8\pi^{2}}{\gym^{2}}$.} and one finds
(we keep only the couplings up to $t_{8}$ and dots in the coefficients of the $\l^{n}$ terms denote higher $t_{2n}$ contributions)
\ba
\la{3.2}
\sF_{0}(\l) &= \frac{t_2 \l }{16 \pi ^2}+\frac{t_4 \l ^2}{128 \pi ^4}+\frac{5 t_6 \l \
^3}{4096 \pi ^6}+\frac{7 t_8 \l ^4}{32768 \pi ^8}+\cdots,  \notag \\ 
\sF_{1}(\l) &=  -\frac{t_2^2 \l ^2}{256 \pi ^4}-\frac{(t_2 t_4) \l ^3}{512 \pi 
^6}+\frac{(-18 t_4^2-30 t_2 t_6) \l ^4}{65536 \pi ^8}+\frac{(-720 t_4 
t_6-560 t_2 t_8) \l ^5}{5242880 \pi ^{10}}\lp
+\frac{(-900 t_6^2-1680 t_4 
t_8+\cdots) \l ^6}{50331648 \pi ^{12}}-\frac{15 (40 t_6 
t_8+\cdots) \l ^7}{67108864 \pi ^{14}}-\frac{175 (14 t_8^2+\cdots) \l ^8}{2147483648 \pi ^{16}}+\cdots, \notag \\
\sF_{2}(\l) &=  \frac{t_4 \l ^2}{256 \pi ^4}+\frac{(4 t_2^3+30 t_6) \l ^3}{12288 \pi 
^6}+\frac{(24 t_2^2 t_4+70 t_8) \l ^4}{65536 \pi ^8}+\frac{(720 t_2 
t_4^2+600 t_2^2 t_6+\cdots) \l ^5}{5242880 \pi ^{10}}\lp
+\frac{(864 
t_4^3+4320 t_2 t_4 t_6+1680 t_2^2 t_8) \l ^6}{50331648 \pi 
^{12}}+\frac{(7560 t_4^2 t_6-21 t_2 (-300 t_6^2-560 t_4 t_8)+\cdots) \l ^7}{469762048 \pi ^{14}}\lp
+\frac{(10800 t_4 
t_6^2+10080 t_4^2 t_8+420 t_2 (40 t_6 t_8+\cdots)) \l 
^8}{2147483648 \pi ^{16}}+\cdots\ .
\ea
Of course, by replacing in (\ref{3.2}) the couplings in the sum in (\ref{2.13}) one gets back (\ref{3.1}).
We can write the cluster expansion of $\Delta F_{N}(\bm t)$ as 
\ba
\la{3.3}
\Delta \sF_{N}(\bm t) &= \sum_{n=1}^{\infty} \bigg(\frac{\l}{8\pi^{2}}\bigg)^{n}\,C_{n}(N)\,t_{2n} 
-\frac{1}{2}\sum_{n,m=1}^{\infty} \bigg(\frac{\l}{8\pi^{2}}\bigg)^{n+m}\,C_{n,m}(N)\, t_{2n}\, t_{2m} \lp
 +\frac{1}{3!}\sum_{n,m,k=1}^{\infty} \bigg(\frac{\l}{8\pi^{2}}\bigg)^{n+m+k}\,C_{n,m,k}(N)\, t_{2n}\,t_{2m}\,t_{2k}+\cdots\ ,
 \ea
 where
 \be
\la{3.4}
 C_{n}(N) = \vev{ \tr \big(\tfrac{A}{\sqrt N}\big)^{2n}}, \qquad C_{n,m}(N) = \vev{ \tr \big(\tfrac{A}{\sqrt N}\big)^{2n}\big(\tfrac{A}{\sqrt N}\big)^{2m}}_{c}, \dots.
 \ee
All dependence on $N$ is captured by the connected correlators $C_{n,m,\dots}(N)$ and their $1/N$ expansion provides the $1/N$ expansion of $\Delta \sF$.
We have the following explicit expressions for the Gaussian correlators in (\ref{3.4})
 \footnote{ 
$\vev{ABC}_{c} = 
\vev{ABC}-\vev{A}\, \vev{BC}-\vev{B}\, \vev{AC}-\vev{C}\, \vev{AB}+2\vev{A}\, 
\vev{B}\, \vev{C}$, {\em etc.}}
\ba
\la{3.5}
C_{n}(N) &= N\,\frac{2^{n}\Gamma(n+\frac{1}{2})}{\sqrt \pi\,\Gamma(n+2)}\bigg[1+\frac{1}{N^{2}}\frac{n(n^{2}-1)}{12}+\frac{1}{N^{4}}\frac{n(n^{2}-1)(n-2)(n-3)(5n-2)}{1440}+\cdots\bigg], \\
\nonumber 
C_{n,m}(N) &=  \frac{2^{n+m}\Gamma(n+\frac{1}{2})\Gamma(m+\frac{1}{2})}{\pi\,(n+m)\,\Gamma(n)\Gamma(m)} \bigg[1+\frac{1}{N^{2}}\frac{(n+m)(1-2n-2m+n^{2}+nm+m^{2})}{12}+\cdots\bigg], \\
\nonumber 
C_{n,m,k}(N) &=  
\frac{1}{N}\,\frac{2^{n+m+k}\Gamma(n+\frac{1}{2})\Gamma(m+\frac{1}{2})\Gamma(k+\frac{1}{2})}
{\pi^{3/2}\,\Gamma(n)\Gamma(m)\Gamma(k)}\,\bigg[1+\frac{1}{N^{2}}c_{n,m,k}+\cdots\bigg], 
\ea
with
\ba
c_{n,m,k} &= -\frac{1}{6}+\frac{5}{12}(n+m+k)-\frac{1}{6}[2(n^{2}+m^{2}+k^{2})+3(nm+nk+mk)] \lp
+\frac{1}{12}[n^{3}+m^{3}+k^{3}+2(k^{2}m+km^{2}+k^{2}n+kn^{2}+mn^{2}+m^{2}n)+2nmk.
\ea
From these results we can obtain the explicit expression of $\sF_{0}$, $\sF_{1}$, $\sF_{2}$. The linear in $N$ term is  
\ba
\sF_{0}(\l, \bm t) &= \sum_{n=1}^{\infty} \bigg(\frac{\l}{4\pi^{2}}\bigg)^{n}\,\,\frac{\Gamma(n+\frac{1}{2})}{\sqrt \pi\,\Gamma(n+2)}\,t_{2n},
\ea
and can be written in the convenient form 
\be
\la{3.8}
\sF_{0}(\l) = -\oint\frac{dx}{2\pi i}(x-x^{-1})L\left(\frac{\sql}{4\pi}(x+x^{-1})\right),
\ee
due to the relation
\be
 -\oint\frac{dx}{2\pi i}(x-x^{-1})\left(\frac{\sql}{4\pi}(x+x^{-1})\right)^{2n} = \bigg(\frac{\l}{4\pi^{2}}\bigg)^{n}\,\,\frac{\Gamma(n+\frac{1}{2})}{\sqrt \pi\,\Gamma(n+2)}.
\ee
The next correction is 
\ba
\sF_{1}(\l, \bm t) &= -\frac{1}{2}\sum_{n,m=1}^{\infty} \bigg(\frac{\l}{4\pi^{2}}\bigg)^{n+m}\, \frac{\Gamma(n+\frac{1}{2})\Gamma(m+\frac{1}{2})}{\pi\,(n+m)\,\Gamma(n)\Gamma(m)}\, t_{2n}\, t_{2m},
 \ea
 that implies
 \ba
 \l\del \sF_{1}(\l, \bm t) &= -\frac{1}{2}\sum_{n,m=1}^{\infty} \bigg(\frac{\l}{4\pi^{2}}\bigg)^{n+m}\, \frac{\Gamma(n+\frac{1}{2})\Gamma(m+\frac{1}{2})}{\pi\,\Gamma(n)\Gamma(m)}\, t_{2n}\, t_{2m}\lp
 = -\frac{1}{2}\bigg[\sum_{n}^{\infty} \bigg(\frac{\l}{4\pi^{2}}\bigg)^{n}\,n(n+1) \frac{\Gamma(n+\frac{1}{2})}{\sqrt\pi\,\Gamma(n+2)}\, t_{2n}\bigg]^{2} = -\frac{1}{2}[\l\del^{2}[\l\sF_{0}(\l, \bm{t})]]^{2},
 \ea
 or (omitting arguments)
 \be
 \la{3.12}
\del\sF_{1}= -\frac{1}{2}\l\,[(\l\sF_{0})'']^{2}.
 \ee
 Finally, at the next order in the $1/N$ expansion, we obtain 
 \ba
 \la{3.13}
\sF_{2}(\l, \bm t)  &= \sum_{n=1}^{\infty} \bigg(\frac{\l}{4\pi^{2}}\bigg)^{n}\,\frac{\Gamma(n+\frac{1}{2})}{\sqrt \pi\,\Gamma(n+2)}\frac{n(n^{2}-1)}{12}\,t_{2n} \lp
 +\frac{1}{6}\sum_{n,m,k=1}^{\infty} \bigg(\frac{\l}{4\pi^{2}}\bigg)^{n+m+k}\,\frac{\Gamma(n+\frac{1}{2})\Gamma(m+\frac{1}{2})\Gamma(k+\frac{1}{2})}
{\pi^{3/2}\,\Gamma(n)\Gamma(m)\Gamma(k)}\, t_{2n}\,t_{2m}\,t_{2k},
\ea
that we can write
\be
\la{3.14}
\sF_{2} = \frac{1}{12}\l\del ((\l\del)^{2}-1)\sF_{0}+\frac{1}{6}[\l\del^{2}(\l\sF_{0})]^{3}.
 \ee
 It is clear that we can continue generating differential relations by extending the expansions (\ref{3.5}).
 In the next section, we present a general formalism to derive systematically the above  relations by
 exploiting the integrable hierarchy governing the matrix model.

\section{Integrable hierarchies and hermitian 1-matrix models}
\la{sec:hier}

Let us briefly recall some known facts about orthogonal polynomials and their role in the evaluation of single trace matrix model partition functions
\cite{Gerasimov:1990is}.
Let us introduce the measure associated with the potential $W$ in (\ref{2.10})
\be
d\mu(\bm t) = e^{-W(x; \bm t)}\frac{dx}{2\pi}.
\ee
We consider monic polynomials $P_{n}(x; \bm t) = x^{n}+\cdots$ orthogonal with respect to  $d\mu(\bm t) $
\be
\la{4.2}
\vev{n|m} = \int d\mu(\bm t)\, P_{n}(x; \bm t) P_{m}(x; \bm t) = h_{n}(\bm t)\delta_{nm}, \qquad h_{n}(\bm t) \equiv e^{-f_{n}(\bm t)}.
\ee
The partition function in (\ref{2.10}) and the associated free energy obey
\be
Z_{N}(\bm t) = \prod_{n=0}^{N-1}h_{n}(\bm t), \qquad \to \qquad F_{N}(\bm t) =\sum_{n=0}^{N-1}f_{n}(\bm t).
\ee
We can write 
\be
\la{4.4}
f_{N}(\bm t) = \mc D_{+}F_{N}(\bm t), \qquad \text{and}\qquad f_{N}(\bm t)-f_{N-1}(\bm t) = \mc D^{2}F_{N}(\bm t),
\ee
where we have introduced the (commuting) forward/backward difference operators
\be
\mc D_{+}X_{N} = X_{N+1}-X_{N},\qquad \mc D_{-}X_{N} = X_{N}-X_{N-1},
\ee
and the second order operator
\be
\mc D^{2} = \mc D_{+}\mc D_{-}, \qquad \mc D^{2}X_{N} = X_{N+1}-2X_{N}+X_{N-1}.
\ee
Let us also introduce the auxiliary quantities
\be
\la{4.7}
r_{n}(\bm t) = e^{-f_{n}(\bm t)+f_{n-1}(\bm t)} = e^{-\mc D_{+}f_{n-1}(\bm t)}.
\ee
The \underline{first equation} in the Toda hierarchy associated with this system reads
\ba
-\frac{\partial^{2}f_{n}}{\partial t_{1}^{2}} &=r_{n+1}-r_{n} = e^{-f_{n+1}+f_{n}}-e^{-f_{n}+f_{n-1}}.
\ea
Summing over $n$, it implies 
\ba
\la{4.9}
 \frac{\partial^{2}F_{N}(\bm t)}{\partial t_{1}^{2}} &= \sum_{n=0}^{N-1}\frac{\partial^{2}}{\partial t_{1}^{2}}f_{n} = -e^{-f_{N}+f_{N-1}} = -e^{-\mc D^{2}F_{N}(\bm t)}.
 \ea
 
\paragraph{Volterra reduction}

For an even potential $W$, with only $t_{2n}$ couplings, the Toda hierarchy reduces to the Volterra hierarchy, see Appendix \ref{app:volt}. In this case, the first equation
of the hierarchy reads
\be
\la{4.10}
\frac{\partial f_{n}}{\partial t_{2}} = r_{n+1}+r_{n}.
\ee
In terms of the free energy, this relation implies, \cf (\ref{4.4})  and (\ref{4.7})  with $n=N$, 
\be
\la{4.11}
\frac{\partial}{\partial t_{2}}\mc D_{+}F_{N}(\bm t) = e^{-\mc D^{2}F_{N}(\bm t)}+e^{-\mc D^{2}F_{N+1}(\bm t)}.
\ee

\paragraph{Remark:} The potential for the D3-D5 system, \cf  (\ref{2.10}) and  (\ref{2.12}), is even. Nevertheless, if one is interested
in gauge group $SU(N)$ instead of $U(N)$, it is natural to represent the traceless constraint by introducing an auxiliary coupling $t_{1}$.
For this reason, in the later section \ref{sec:sun} we will need some consequences of the Toda equation (\ref{4.9}). In particular, we will need 
the multi-trace defect one-point functions
$\vevD{(\tr M)^{n}}$ whose computation is discussed in Appendix \ref{app:t1}.

\section{All order $1/N$ expansion of the free energy in the $U(N)$ model from the Volterra hierarchy}
\la{sec:free-exp}

In our model, the free energy has also a dependence on $\gym$, but it may be linked to the dependence on $t_{2}$ since 
\ba
 & F_{N}(\gym, \bm t) =  -\log\int_{\mathfrak{u}(N)} \mc DA\, e^{-\tr A^{2}-\sum_{n\ge 1}t_{2n}\left(\frac{\gym^{2}}{8\pi^{2}}\right)^{n}\tr A^{2n}}\lp
=  -\frac{N^{2}}{2}\log\frac{8\pi^{2}}{\gym^{2}}-\log \int_{\mathfrak{u}(N)} \mc DA\, e^{-\left(\frac{8\pi^{2}}{\gym^{2}}+t_{2}\right)\tr A^{2}-\sum_{n\ge 2}t_{2n}\tr A^{2n}} \lp
= -\frac{N^{2}}{2}\log\frac{8\pi^{2}}{\gym^{2}}+\widehat{F}_{N}\left(t_{2}+\frac{8\pi^{2}}{\gym^{2}}, t_{3}, \dots\right).
 \ea
 Hence, we can replace $\frac{8\pi^{2}}{\gym^{2}}\partial_{t_{2}}$ by differentiation with respect to $\gym$ as follows
 \ba
 \la{5.2}
\frac{8\pi^{2}}{\gym^{2}} \partial_{t_{2}}F_{N}(\gym, \bm t) &= \frac{8\pi^{2}}{\gym^{2}}\frac{\partial}{\partial\frac{8\pi^{2}}{\gym^{2}}}\widehat{F}_{N}\left(t_{2}+\frac{8\pi^{2}}{\gym^{2}}, t_{3}, \dots\right) 
 = -\frac{1}{2}\gym\frac{\partial}{\partial\gym}[F_{N}(\gym, \bm t)+\frac{N^{2}}{2}\log\frac{8\pi^{2}}{\gym^{2}}]\lp
 = -\frac{1}{2}\gym\frac{\partial}{\partial\gym}F_{N}(\gym, \bm t)+\frac{N^{2}}{2}.
 \ea
Besides
\be
\la{5.3}
F_{N}= \Delta F_{N} +F_{N}^{\rm Gaussian} = \Delta F_{N}-\log\bigg[(2\pi)^{-\frac{N}{2}}\,G(N+1)\,2^{-\frac{N^{2}}{2}}\bigg],
\ee
where $G$ is the Barnes $G$-function.
Plugging (\ref{5.2}) and (\ref{5.3}) into (\ref{4.11}) gives \footnote{Notice that 
if we apply $\mc D_{-}$ to this we get the more symmetric form 
\be
\notag 
-\frac{1}{2}\gym\partial_{g}\mc D^{2}\Delta F_{N}(\gym, \bm t)+1 = \frac{1}{2}\bigg[(N+1)\,e^{-\mc D^{2}\Delta F_{N+1}(\gym, \bm t)}-(N-1)\,e^{-\mc D^{2}\Delta F_{N-1}(\gym, \bm t)}\bigg].
\ee
}
\be
-\gym\partial_{g}\mc D_{+}\Delta F_{N}(\gym, \bm t)+2N+1 =\bigg[N\,e^{-\mc D^{2}\Delta F_{N}(\gym, \bm t)}+(N+1)\,e^{-\mc D^{2}\Delta F_{N+1}(\gym, \bm t)}\bigg]
\ee
Finally, we turn this equation into a differential-difference equation in $\l$ 
 \be
 \la{5.5}
 2N+1-2\l\del\mc D\Delta F_{N}(\l) = N\, e^{-\mc D^{2}\Delta F_{N}(\l)} +(N+1) e^{-\mc D^{2}\Delta F_{N+1}(\l)}.
 \ee
  where 
\ba
 \la{5.6}
 \mc D^{2}\Delta F_{N} &= \mc D^{2}\Delta F_{N}(\gym^{2}N)  
= \Delta F_{N+1}(\l\tfrac{N+1}{N})-2\Delta F_{N}(\l)+\Delta F_{N-1}(\l\tfrac{N-1}{N}),
 \ea
 and similarly
\be
\la{5.7}
\mc D^{2}\Delta F_{N+1} = 
\Delta F_{N+ 2}(\l\tfrac{N+ 2}{N})-2\Delta F_{N+ 1}(\l\tfrac{N+ 1}{N})+\Delta F_{N}(\l).
\ee
As we now show, the master equation (\ref{5.5}) encodes the differential relations (\ref{3.12}) and (\ref{3.14}) as special cases and generalizes them to higher order.
 
%
%
%

\paragraph{Differential relations}

Let us plug the expansion  (\ref{2.17}) in (\ref{5.5}).  The first non-trivial relation is 
\be
\la{5.8}
\sF_{1}' = -2 \lambda  \sF_0'{}^2-2 \lambda ^2 \sF_0' \sF_0''-\frac{1}{2} \lambda ^3 \sF_0''{}^2,
\ee
which is the same as (\ref{3.12}). At the next order in $1/N$, and using systematically (\ref{5.8}), we obtain 
\be
\sF_{2} = \frac{4}{3} \l ^3 \sF_0'{}^3+\frac{1}{4} \l ^2 \sF_0''+2 \l ^4 
\sF_0'{}^2 \sF_0''+\l ^5 \sF_0' \sF_0''{}^2+\frac{1}{6} \l ^6 
\sF_0''{}^3+\frac{1}{12} \l ^3 \sF_0{}^{(3)},
\ee
which agrees with (\ref{3.14}). The next coefficient functions turn out to be all expressible by algebraic differential operators acting on $\sF_{0}$. For instance, we find 
\ba
\sF_{3} &= -2 \l ^4 \sF_0'{}^4-\l ^3 \sF_0' \sF_0''-8 \l ^5 \sF_0'{}^3 
\sF_0''-\frac{7}{8} \l ^4 \sF_0''{}^2-9 \l ^6 \sF_0'{}^2 
\sF_0''{}^2-4 \l ^7 \sF_0' \sF_0''{}^3-\frac{5}{8} \l ^8 \sF_0''{}^4 \lp
-\l ^4 \sF_0' \sF_0{}^{(3)}-\frac{4}{3} \l ^6 \sF_0'{}^3 
\sF_0{}^{(3)}-\frac{3}{4} \l ^5 \sF_0'' \sF_0{}^{(3)}-2 \l ^7 
\sF_0'{}^2 \sF_0'' \sF_0{}^{(3)}-\l ^8 \sF_0' \sF_0''{}^2 
\sF_0{}^{(3)}-\frac{1}{6} \l ^9 \sF_0''{}^3 
\sF_0{}^{(3)}\lp
-\frac{1}{24} \l ^6 \sF_0{}^{(3)}{}^2-\frac{1}{6} \l ^5 
\sF_0' \sF_0{}^{(4)}-\frac{1}{12} \l ^6 \sF_0'' \sF_0{}^{(4)}.
\ea
 and
 \ba
 \sF_{4} &= \frac{16}{5} \l ^5 \sF_0'{}^5+3 \l ^4 \sF_0'{}^2 \sF_0''+24 \l ^6 
\sF_0'{}^4 \sF_0''+9 \l ^5 \sF_0' \sF_0''{}^2+52 \l ^7 \sF_0'{}^3 
\sF_0''{}^2+\frac{9}{2} \l ^6 \sF_0''{}^3+46 \l ^8 \sF_0'{}^2 
\sF_0''{}^3\lp
+18 \l ^9 \sF_0' \sF_0''{}^4+\frac{13}{5} \l ^{10} 
\sF_0''{}^5+5 \l ^5 \sF_0'{}^2 \sF_0{}^{(3)}+8 \l ^7 \sF_0'{}^4 
\sF_0{}^{(3)}+13 \l ^6 \sF_0' \sF_0'' \sF_0{}^{(3)}+24 \l ^8 
\sF_0'{}^3 \sF_0'' \sF_0{}^{(3)}\lp
+6 \l ^7 \sF_0''{}^2 \sF_0{}^{(3)}+24 
\l ^9 \sF_0'{}^2 \sF_0''{}^2 \sF_0{}^{(3)}+10 \l ^{10} \sF_0' 
\sF_0''{}^3 \sF_0{}^{(3)}+\frac{3}{2} \l ^{11} \sF_0''{}^4 
\sF_0{}^{(3)}+2 \l ^7 \sF_0' \sF_0{}^{(3)}{}^2+\frac{4}{3} \l ^9 
\sF_0'{}^3 \sF_0{}^{(3)}{}^2\lp
+\frac{5}{4} \l ^8 \sF_0'' 
\sF_0{}^{(3)}{}^2+2 \l ^{10} \sF_0'{}^2 \sF_0'' \sF_0{}^{(3)}{}^2+\l 
^{11} \sF_0' \sF_0''{}^2 \sF_0{}^{(3)}{}^2+\frac{1}{6} \l ^{12} 
\sF_0''{}^3 \sF_0{}^{(3)}{}^2+\frac{1}{36} \l ^9 
\sF_0{}^{(3)}{}^3+\frac{1}{16} \l ^4 \sF_0{}^{(4)}\lp
+\frac{11}{6} \l ^6 
\sF_0'{}^2 \sF_0{}^{(4)}+\frac{2}{3} \l ^8 \sF_0'{}^4 
\sF_0{}^{(4)}+\frac{17}{6} \l ^7 \sF_0' \sF_0'' 
\sF_0{}^{(4)}+\frac{4}{3} \l ^9 \sF_0'{}^3 \sF_0'' 
\sF_0{}^{(4)}+\frac{23}{24} \l ^8 \sF_0''{}^2 \sF_0{}^{(4)}\lp
+\l ^{10} 
\sF_0'{}^2 \sF_0''{}^2 \sF_0{}^{(4)}+\frac{1}{3} \l ^{11} \sF_0' 
\sF_0''{}^3 \sF_0{}^{(4)}+\frac{1}{24} \l ^{12} \sF_0''{}^4 
\sF_0{}^{(4)}+\frac{1}{3} \l ^8 \sF_0' \sF_0{}^{(3)} 
\sF_0{}^{(4)}+\frac{1}{6} \l ^9 \sF_0'' \sF_0{}^{(3)} 
\sF_0{}^{(4)}+\frac{1}{30} \l ^5 \sF_0{}^{(5)}\lp
+\frac{1}{6} \l ^7 
\sF_0'{}^2 \sF_0{}^{(5)}+\frac{1}{6} \l ^8 \sF_0' \sF_0'' 
\sF_0{}^{(5)}+\frac{1}{24} \l ^9 \sF_0''{}^2 
\sF_0{}^{(5)}+\frac{1}{288} \l ^6 \sF_0{}^{(6)}.
\ea
Remarkably, it turns out that these quantities take a simplified form in terms of 
\be
\la{5.12}
Z(\l) = \l\,(\l\,\sF_{0})''.
\ee
Explicit expressions are collected in Appendix \ref{app:ZZZ}.

\subsection{Strong coupling expansion}

Let us now specialize to the D3-D5 potential. We begin with the leading order (\ref{3.8})
\ba
\la{5.13}
\sF_{0}(\l) &= -\oint\frac{dx}{2\pi i }(x-x^{-1})\log\cosh\bigg[\frac{\sql}{4}(x+x^{-1})\bigg] = \frac{4}{\pi}\int_{0}^{\pi/2}d\theta\,\sin^{2}\theta\,\log\cosh\bigg(\frac{\sql}{2}\cos\theta\bigg) \lp
= \frac{4}{\pi}\int_{0}^{1}dt\, \sqrt{1-t^{2}}\log\cosh\frac{t\sql}{2} = \frac{2}{3\pi}\sql-\log 2+\frac{4}{\pi}\int_{0}^{1}dt\, \sqrt{1-t^{2}}\log(1+e^{-t\sql}).
\ea
The last integral is subleading but not exponentially suppressed due to the point $t=0$ where exponential suppression does not hold.
From the results in Appendix \ref{app:asymp} we have 
the asymptotic expansion \footnote{Notice that the infinite sum in (\ref{5.14}) has a term $\sim\sql$ that is dominant with respect to the constant $-\log 2$. That constant term is somehow separated since it 
arises as an integration constant.}
\be
\la{5.14}
\sF_{0}(\l) = -\log 2-\frac{2}{\pi^{2}}\sum_{k=0}^{\infty}\frac{2^{2k-1}-1}{2k-1}\Gamma\left(k-\frac{3}{2}\right)\Gamma\left(k+\frac{1}{2}\right)\zeta(2k)\frac{1}{\l^{k-1/2}}.
\ee
The first terms are explicitly
\be
\la{5.15}
\sF_{0}(\l) = \frac{2}{3\pi}\sql-\log 2+\frac{\pi}{3}\frac{1}{\sql}-\frac{7\pi^{3}}{180}\frac{1}{\l^{3/2}}-\frac{31\pi^{5}}{2520}\frac{1}{\l^{5/2}}-\frac{127\pi^{7}}{6720}\frac{1}{\l^{7/2}}+\cdots.
\ee
Using the differential relations for $\sF_{n}(\l)$ we obtain the following non vanishing terms at large coupling $\l$ \footnote{
We remark the presence of a $\log \l$ term in $\sF_{1}(\l)$, \ie at the next-to-leading term in the $1/N$ expansion. It would be interesting to understand whether 
it may admit an interpretation in terms of dCFT conformal anomalies, see for instance \cite{Herzog:2020wlo}, or it is instead independent and a feature of the 
non-Gaussian matrix model.
}
\ba
\la{5.16}
\sF_{1}(\l) &= -\frac{\l}{8\pi^{2}}+\frac{1}{24}\log\l+\mc O(1/\sql), \notag \\
\sF_{2}(\l) &= \frac{\l^{3/2}}{48\pi^{3}}-\frac{\sql}{32\pi}+\mc O(1/\sql),  \notag \\
\sF_{3}(\l) &= -\frac{\l^{2}}{384\pi^{4}}+\frac{5\l^{3/2}}{18432\pi^{3}}+\frac{1}{240}+\mc O(1/\sql), \\
\sF_{4}(\l) &= \frac{\l^{5/2}}{5120\pi^{5}}+\frac{5\l^{3/2}}{18432\pi^{3}}+\frac{79\sql}{368640\pi}+\mc O(1/\sql),  \notag \\
\sF_{5}(\l) &= -\frac{\l^{2}}{12288\pi^{4}}-\frac{11\l}{23040\pi^{2}}-\frac{1847}{1161216}+\mc O(1/\sql), \notag \\
\sF_{6}(\l) &= -\frac{\l^{7/2}}{688128\pi^{7}}-\frac{7\l^{5/2}}{983040\pi^{5}}+\frac{55\l^{3/2}}{786432\pi^{3}}+\frac{122947\sql}{212336640\pi}+\mc O(1/\sql).\notag 
\ea
Keeping the leading terms at each order in the $1/N$ expansion, these results suggest the strong coupling scaling form 
\be
\la{5.17}
\Delta F(\l, N) = \l\,f\left(\frac{\l}{N^{2}}\right)+\cdots,
\ee
where dots are subleading terms at large $N$ with fixed $\l/N^{2}$. 
Comparing with the above gives
\be
\la{5.18}
f(x) = \frac{2}{3 \pi  \sqrt{x}}-\frac{1}{8 \pi ^2}+\frac{\sqrt{x}}{48 \pi 
^3}-\frac{x}{384 \pi ^2}+\frac{x^{3/2}}{5120 \pi 
^5}-\frac{x^{5/2}}{688128 \pi ^7}+\cdots\ .
\ee
Indeed, inserting the Ansatz (\ref{5.17}) into (\ref{5.5}) gives the non-linear differential equation 
\be
\la{5.19}
1-e^{-x^{3}f''}-x(f-xf'-x^{2}f'')=0.
\ee
The solution is obtained in Appendix \ref{app:diff} and  reads
\be
\la{5.20}
f(x) = \frac{1}{384\pi^{2}}\bigg[-48-x+8\bigg(5+\frac{x}{8\pi^{2}}\bigg)\,\sqrt{1+\frac{16\pi^{2}}{x}}+\frac{192\pi^{2}}{x}\,\text{arccosh}\bigg(1+\frac{x}{8\pi^{2}}\bigg)\bigg].
\ee

\section{Free energy in the $SU(N)$ model at order $1/N^{3}$}
\la{sec:free-sun}

If the gauge group is $SU(N)$ we need to consider an extra traceless condition in the matrix model. This is a non-trivial modification, although irrelevant at leading large $N$.
It is due to the fact that the $U(1)$ degrees of freedom in the two $\mc N=4$ SYMs do not decouple due to interaction with the interface.  
We remark that it would be interesting to clarify the string counterpart of this fact.

The traceless condition complicates the application of the integrable hierarchies at least at practical level. 
Anyway, we can directly determine the $1/N$ expansion of the defect expectation value (our free energy)
by the methods in section \ref{sec:free}. Denoting by a tilde the quantities in the $SU(N)$ case, the expansions (\ref{3.1}) are modified into  
\ba
\tilde{\sF}_{0}(\l) &= \sF_{0}(\l), \notag \\
\tilde{\sF}_{1}(\l) &= \sF_{1}(\l), \notag \\
\tilde{\sF}_{2}(\l) &=-\frac{\l }{32}+\frac{5 \l ^2}{3072}-\frac{5 \l 
^3}{73728}-\frac{\l ^4}{2359296}+\frac{191 \l 
^5}{283115520}-\frac{4259 \l ^6}{33973862400}+\cdots ,
\ea
with deviations in all $\tilde{\sF}_{n\ge 2}$. 
Inspection suggests the simple relation 
\be
\la{6.2}
\tilde{\sF}_{2} = \sF_{2}-\frac{\l}{2}\,(\l \sF_{0})''.
\ee
To prove (\ref{6.2})  and generalize it, we start again from the connected correlators in (\ref{3.5}) that, in $SU(N)$ case, read, \cf Appendix \ref{app:Cn}, 
\ba
\la{6.3}
\nonumber
C^{SU(N)}_{n}(N) &= N\,\frac{2^{n}\Gamma(n+\frac{1}{2})}{\sqrt \pi\,\Gamma(n+2)}\bigg[1+\frac{1}{N^{2}}\frac{n(n+1)(n-7)}{12}
+\frac{1}{N^{4}}\frac{n(n^{2}-1)(5n^{3}-87n^{2}+340n-12)}{1440}+\cdots\bigg], \\
\nonumber 
C^{SU(N)}_{n,m}(N) &=  \frac{2^{n+m}\Gamma(n+\frac{1}{2})\Gamma(m+\frac{1}{2})}{\pi\,(n+m)\,\Gamma(n)\Gamma(m)} \bigg[
1+\frac{1}{N^{2}}\frac{(n+m)(7-8n-8m+n^{2}+nm+m^{2})}{12}+\cdots
\bigg], \\
C^{SU(N)}_{n,m,k}(N) &=  
\frac{1}{N}\,\frac{2^{n+m+k}\Gamma(n+\frac{1}{2})\Gamma(m+\frac{1}{2})\Gamma(k+\frac{1}{2})}
{\pi^{3/2}\,\Gamma(n)\Gamma(m)\Gamma(k)}\,\bigg[1+\frac{1}{N^{2}}\tilde c_{n,m,k}+\cdots\bigg],
\ea
where
\ba
\tilde c_{n,m,k} &= -\frac{7}{6}+\frac{23}{12}(n+m+k)-\frac{1}{6}[5(n^{2}+m^{2}+k^{2})+9(nm+nk+mk)] \lp
+\frac{1}{12}[n^{3}+m^{3}+k^{3}+2(k^{2}m+km^{2}+k^{2}n+kn^{2}+mn^{2}+m^{2}n)+2nmk.
\ea
Notice that defining $\Delta C = C^{SU(N)}-C^{U(N)}$ we have, \cf (\ref{3.5}),
\ba
\Delta C_{n}(N) &= N\,\frac{2^{n}\Gamma(n+\frac{1}{2})}{\sqrt \pi\,\Gamma(n+2)}\bigg[-\frac{1}{N^{2}}\frac{n(n+1)}{2}
-\frac{1}{N^{4}}\frac{n^{2}(n^{2}-1)(n-5)}{24}+\cdots\bigg], \\
\nonumber 
\Delta C_{n,m}(N) &=  \frac{2^{n+m}\Gamma(n+\frac{1}{2})\Gamma(m+\frac{1}{2})}{\pi\,(n+m)\,\Gamma(n)\Gamma(m)} \bigg[
-\frac{1}{N^{2}}\frac{(n+m)(n+m-1)}{2}+\cdots
\bigg], \\
\nonumber 
\Delta C_{n,m,k}(N) &=  
\frac{1}{N}\,\frac{2^{n+m+k}\Gamma(n+\frac{1}{2})\Gamma(m+\frac{1}{2})\Gamma(k+\frac{1}{2})}
{\pi^{3/2}\,\Gamma(n)\Gamma(m)\Gamma(k)}\,\bigg[\frac{1}{N^{2}}\frac{(n+m+k-1)(n+m+k-2)}{2}+\cdots\bigg],
\ea
Hence, for instance, (\ref{3.13}) reads 
\ba
\tilde{\sF}_{2}(\l, \bm t)  &= \sum_{n=1}^{\infty} \bigg(\frac{\l}{4\pi^{2}}\bigg)^{n}\,\frac{\Gamma(n+\frac{1}{2})}{\sqrt \pi\,\Gamma(n+2)}\bigg[\frac{n(n^{2}-1)}{12}\blue{-\frac{1}{2}n(n+1)}\bigg]\,t_{2n} \lp
 +\frac{1}{6}\sum_{n,m,k=1}^{\infty} \bigg(\frac{\l}{4\pi^{2}}\bigg)^{n+m+k}\,\frac{\Gamma(n+\frac{1}{2})\Gamma(m+\frac{1}{2})\Gamma(k+\frac{1}{2})}
{\pi^{3/2}\,\Gamma(n)\Gamma(m)\Gamma(k)}\, t_{2n}\,t_{2m}\,t_{2k},
\ea
where the blue term is the shift with respect to (\ref{3.13}) and provides the second term in (\ref{6.2}). 
We can analyze in a similar way the  $1/N^{2}$ term $\tilde{\sF}_{3}$ and the $1/N^{3}$ term $\tilde{\sF}_{4}$ and we obtain
\ba
\la{6.7}
\tilde{\sF}_{3} &= \sF_{3}-\frac{\l^{2}}{2}\,\sF_{1}'' =  \sF_{3}-\frac{\l^{2}}{2}(-2 \lambda  \sF_0'{}^2-2 \lambda ^2 \sF_0' \sF_0''-\frac{1}{2} \lambda ^3 \sF_0''{}^2)', \\
\tilde{\sF}_{4} &= \sF_{4}+\l^{2}\bigg[
\frac{\sF_0'}{3}-\frac{2}{3} \l ^2 \sF_0'{}^3+\frac{5}{12} \l  \sF_0''-4 
\l ^3 \sF_0'{}^2 \sF_0''-\frac{7}{2} 
\l ^4 \sF_0' \sF_0''{}^2-\frac{5}{6}
\l ^5 \sF_0''{}^3  \lp
-\frac{1}{12} \l ^2 
\sF_0^{(3)}-\l ^4 \sF_0'{}^2 
\sF_0^{(3)}-\l ^5 \sF_0' 
\sF_0'' \sF_0^{(3)}-\frac{1}{4} \l ^6 
\sF_0''{}^2 \sF_0^{(3)}-\frac{1}{24} 
\l ^3 \sF_0^{(4)}
\bigg]'.
\ea
We can examine the effect of these shifts at strong coupling using (\ref{5.15}). This gives
\ba
\tilde \sF_{2}-\sF_{2} &= -\frac{\sql}{4\pi}+\cdots, \\
\tilde \sF_{3}-\sF_{3} &= \frac{1}{48}+\cdots, \\
\tilde \sF_{4}-\sF_{4} &= \frac{\l^{3/2}}{384\pi^{3}}-\frac{5\sql}{256\pi}+\cdots.
\ea
Comparing with (\ref{5.16}) we see that the leading term at large $\l$ is not changed and keep being resummed by the 
expression (\ref{5.17}) and (\ref{5.20}).

 \section{Systematic relation between the free energies in the $U(N)$ and $SU(N)$ models}
 \la{sec:sun}

The shifts $\tilde{\sF}_{n}-\sF_{n}$ can be analyzed in a systematic way as follows.
The partition function in the $SU(N)$ matrix model, with insertion of a generic function $f(A)$,  is 
 \ba
Z_{f}^{\rm SU} &= \int \mc DA\ \delta(\tr A) e^{- \tr A^{2}} f(A) = \frac{1}{2\pi}\int d\alpha \int \mc DA e^{-\tr A^{2}+i\alpha\tr A} f(A) \lp
 = \frac{1}{2\pi}\int d\alpha e^{-\frac{N\alpha^{2}}{4}}\int \mc DA e^{-\tr A^{2}} f(A+\tfrac{i\alpha}{2}) 
 = \frac{1}{\pi\sqrt{2 N}}\int d\alpha e^{-\frac{\alpha^{2}}{2}}\int \mc DA e^{-\tr A^{2}} f(A+\tfrac{i\alpha}{\sqrt{2N}}) .
 \ea
 Hence, we can write
 \be
 \la{7.2}
 \vev{f}^{\rm SU}  = \frac{\int \mc DA e^{-\tr A^{2}}\vev{f(A+\frac{i\alpha}{\sqrt{2N}})}_{0}}{\int \mc DA e^{-\tr A^{2}}}, \qquad
 \vev{f(A+\tfrac{i\alpha}{\sqrt{2N}})}_{0} = \frac{1}{\sqrt{2\pi}}\int d\alpha e^{-\frac{\alpha^{2}}{2}}f(A+\tfrac{i\alpha}{\sqrt{2N}}).
 \ee
 Expanding in powers of $\alpha/\sqrt{N}$ and integrating over $\alpha$, we get
 \be
 \vev{f(A+\tfrac{i\alpha}{\sqrt{2N}})}_{0} = f(A)-\frac{1}{4N}f''(A)+\frac{1}{32N^{2}}f^{(4)}(A)+\cdots = \sum_{p=0}^{\infty}\frac{(-1)^{p}}{p!(4N)^{p}} f^{(2p)}(A).
 \ee
 Inserting this in (\ref{7.2}), and integrating by parts in the matrix model, we obtain  ($H_{n}$ are Hermite polynomials), \cf Appendix \ref{app:herm}, 
 \ba
 \la{7.4}
 \vev{f}^{\rm SU}  &=  \sum_{p=0}^{\infty}\frac{(-1)^{p}}{p!(4N)^{p}} \frac{\int \mc DA [e^{\tr A^{2}}\partial_{A}^{2p}e^{-\tr A^{2}}]f(A)e^{-\tr A^{2}}}{\int \mc DA e^{-\tr A^{2}}}
=  \sum_{p=0}^{\infty}\frac{(-1)^{p}}{p!(4N)^{p}} N^{p}\vev{f(A)\,  H_{2p}(\tfrac{\tr A}{\sqrt N})}.
 \ea
Choosing $f$ to be the defect factor, we have
\ba
\vev{D}^{\rm SU}  &=   \vev{D}\bigg[1+\sum_{p=1}^{\infty}\frac{(-1)^{p}}{p!\,4^{p}} \vevD{ H_{2p}(\tfrac{\tr A}{\sqrt N})}\bigg].
\ea
that gives
\ba
\la{7.6}
\Delta\tilde F &=  \Delta F-\log\bigg[1+\sum_{p=1}^{\infty}\frac{(-1)^{p}}{p!\,4^{p}} \vevD{H_{2p}(\tfrac{\tr A}{\sqrt N})}\bigg].
\ea
Expanding  the logarithm of the series in the r.h.s. we find \footnote{One can check that the two contributions start at $1/N$ and $1/N^{2}$ respectively, \ie  no special cancellations
occur.}
\ba
\Delta\tilde F &=  \Delta F +\frac{1}{N}\bigg(\vevD{\Omega_{1,1}}-\frac{N}{2}\bigg)-\frac{1}{2N^{2}}\bigg(\vevD{\Omega_{1,1,1,1}}-\vevD{\Omega_{1,1}}^{2}
-2N\,\vevD{\Omega_{1,1}}+\frac{N^{2}}{2}\bigg)+\cdots
\ea
From the expansions (\ref{C.7})  and (\ref{C.15}) , using also  (\ref{C.11}) , we obtain 
\be
\Delta\tilde{F} = \Delta F +\frac{1}{N}\omega_{1,1}^{(1)}+\frac{1}{N^{2}}\l^{2}\del\left[\frac{1}{\l}(\omega_{1,1}^{(1)})^{2}\right]+\cdots,
\ee
where $\omega_{1,1}^{(1)}$ is given in (\ref{C.8}). One can check that we prove in this way the formulas for the shifts in $\sF_{2}$ and $\sF_{3}$ given in (\ref{6.2}) and (\ref{6.7}).

\section{One-point functions of single trace operators in the $U(N)$ model}
\la{sec:one-un}

The mixing problem for the operators $\mc O_{2n}$ in (\ref{2.6}) has a large $N$ known solution  in term of Chebyshev T-polynomials, \cf Appendix \ref{app:cheb}, 
\be
:\mc O_{2n}: \stackrel{N\to\infty}{=} \frac{1}{2^{2n-1}}\, \tr T_{2n}(M).
\ee
Since we are interested in higher orders in the $1/N$ expansion, we need finite $N$ expressions. These can be worked out systematically by solving the mixing problem. 
For example, we have the first cases
\ba
\la{8.2}
 :\Omega_{2}: &= \Omega_{2}-\frac{N^{2}}{2}, \qquad
 :\Omega_{4}: = \Omega_{4}-2N\,\Omega_{2}-\Omega_{1,1}+\frac{N(2N^{2}+1)}{4}, \\
 :\Omega_{6}: &= \Omega_{6}-3N\,\Omega_{4}-\frac{3}{2}\Omega_{2,2}+\frac{15}{4}(N^{2}+1)\,\Omega_{2}-3\,\Omega_{1,3}+\frac{15N}{4}\,\Omega_{1,1}
 -\frac{5}{8}N^{2}(N^{2}+2) . \notag 
 \ea
 and one finds
 \ba
 \la{8.3}
 \vevD{ :\Omega_{2}: } &= -\frac{N \l }{32}+\frac{(1+3 N+2 N^2) \l ^2}{1536 \
N}-\frac{(3+7 N+6 N^2+2 N^3) \l ^3}{24576 N^2}+\cdots, \\
 \vevD{ :\Omega_{4}: } &= \frac{(3+5 N+6 N^2+N^3) \l ^2}{3072 N}-\frac{(31+75 N+70 N^2+30 
N^3+4 N^4) \l ^3}{122880 N^2}\lp
+\frac{(840+2251 N+2340 N^2+1225 
N^3+330 N^4+34 N^5) \l ^4}{11796480 N^3}+\cdots , \\
\vevD{:\Omega_{6}:} &= -\frac{(120+318 N+300 N^2+145 N^3+30 N^4+2 N^5) \l ^3}{491520 
N^2}\lp
+\frac{(7785+21385 N+23051 N^2+12775 N^3+3885 N^4+595 N^5+34 N^6) 
\l ^4}{55050240 N^3}+\cdots .
\ea
Going to large $N$, and recalling (\ref{2.22}), we obtain the expansion (\ref{2.19}) with following coefficients $\OO_{n}^{(k)}$. For $\OO_{2}$ 
\ba
\la{8.6}
\OO_{2}^{(0)}(\l) &= \frac{\l}{8\pi^{2}}\,\bigg(-\frac{\l }{32}+\frac{\l ^2}{768}-\frac{\l 
^3}{12288}+\frac{17 \l ^4}{2949120}+\cdots\bigg), \\
\OO_{2}^{(1)}(\l) &=  \frac{\l}{8\pi^{2}}\,\bigg(
\frac{\l ^2}{512}-\frac{\l ^3}{4096}+\frac{11 
\l ^4}{393216}-\frac{29 \l ^5}{9437184}+\cdots\bigg), \notag \\
\OO_{2}^{(2)}(\l) &= \frac{\l}{8\pi^{2}}\,\bigg(
\frac{\l 
^2}{1536}-\frac{7 \l ^3}{24576}+\frac{35 \l 
^4}{589824}-\frac{563 \l ^5}{56623104}+\cdots\bigg). \notag 
\ea
For $\OO_{4}$ 
\ba
\la{8.7}
\OO_{4}^{(0)}(\l) &=\left(\frac{\l}{8\pi^{2}}\right)^{2}\,\bigg(
\frac{\l ^2}{3072}-\frac{\l ^3}{30720}+\frac{17 \l 
^4}{5898240}-\frac{31 \l ^5}{123863040}+\cdots\bigg), \\
\OO_{4}^{(1)}(\l) &= \left(\frac{\l}{8\pi^{2}}\right)^{2}\,\bigg(
\frac{\l 
^2}{512}-\frac{\l ^3}{4096}+\frac{11 \l 
^4}{393216}
-\frac{29 \l ^5}{9437184}+\cdots\bigg), \notag \\
\OO_{4}^{(2)}(\l) &= \left(\frac{\l}{8\pi^{2}}\right)^{2}\,\bigg(
\frac{5 \l 
^2}{3072}-\frac{7 \l ^3}{12288}+\frac{245 \l 
^4}{2359296}-\frac{563 \l ^5}{35389440}+\cdots\bigg).\notag
\ea
Finally, for $\OO_{6}$
\ba
\la{8.8}
\OO_{6}^{(0)}(\l) &=\left(\frac{\l}{8\pi^{2}}\right)^{3}\,\bigg(
-\frac{\l ^3}{245760}+\frac{17 \l ^4}{27525120}-\frac{31 
\l ^5}{440401920}+\frac{691 \l ^6}{95126814720}
+\cdots\bigg), \\
\OO_{6}^{(1)}(\l) &= \left(\frac{\l}{8\pi^{2}}\right)^{3}\,\bigg(
-\frac{\l ^3}{16384}+\frac{17 \l ^4}{1572864}-\frac{91 
\l ^5}{62914560}+\frac{35 \l ^6}{201326592}
+\cdots\bigg), \notag \\
\OO_{6}^{(2)}(\l) &= \left(\frac{\l}{8\pi^{2}}\right)^{2}\,\bigg(
-\frac{29 \l ^3}{98304}+\frac{37 \l ^4}{524288}-\frac{4549 
\l ^5}{377487360}+\frac{56443 \l ^6}{31708938240}
+\cdots\bigg).\notag
\ea
The planar terms of these expansions have been computed  at all orders in \cite{Komatsu:2020sup} and read (for even $n\ge 2$)
\be
\la{8.9}
\OO^{(0)}_{n}(\l) =-n\,
\left(\frac{\l}{16\pi^{2}}\right)^{n/2}
\oint\frac{dx}{2\pi\,i}\frac{1}{x^{n+1}}\log\cosh\bigg[\frac{\sqrt\l}{4}(x+x^{-1})\bigg].
\ee
For a generic defect function $L$ this clearly generalizes to 
\ba
\la{8.10}
\OO^{(0)}_{n}(\l) &=   -n\,\left(\frac{\l}{16\pi^{2}}\right)^{n/2}\oint \frac{dx}{2\pi i}\frac{1}{x^{n+1}}\, L\left(\frac{\sql}{4\pi}(x+x^{-1})\right),
\ea
as derived  in Appendix \ref{app:largeN} by  a slightly different approach compared with 
\cite{Komatsu:2020sup,Wang:2020seq}. In particular, we bypass the resolvent construction and simply solve the $1/N$
perturbation equation for the eigenvalue densities.   

Remarkably, from (\ref{8.10}) we find that all even $\OO^{(0)}_{2n}(\l)$ are  related by an integro-differential relation
\ba
\la{8.11}
\del\OO_{2n}^{(0)}(\l) = \frac{1}{16\pi^{2}}\,\frac{2n}{2n-2}\,\l^{2n-1}\frac{d}{d\l}\bigg[\frac{1}{\l^{2n-2}}\OO^{(0)}_{2n-2}(\l)\bigg].
\ea
To prove this, we notice that the r.h.s. differs from the l.h.s. by  the integral of a non-trivial total derivative:
\ba
& \del\OO_{2n}^{(0)}(\l) - \frac{1}{16\pi^{2}}\,\frac{2n}{2n-2}\,\l^{2n-1}\frac{d}{d\l}\bigg[\frac{1}{\l^{2n-2}}\OO^{(0)}_{2n-2}(\l)\bigg]\lp
=\frac{n}{16\pi^{2}}\left(\frac{\l}{16\pi^{2}}\right)^{n-1}\oint \frac{dx}{2\pi i}\frac{\partial}{\partial x}\bigg[\frac{1+x^{2}}{x^{2n}}L\left(\frac{\sql}{4\pi}(x+x^{-1})\right)\bigg]=0.
\ea

\section{$1/N$ expansion of one-point functions in the $U(N)$ model from the Volterra hierarchy}
\la{sec:one-volterra}

The $1/N$ expansion of one-point functions with gauge group $U(N)$ can be obtained again by exploiting the Volterra hierarchy
and, in the special case of $\OO_{2}$, a special relation.

\subsection*{Results for $\OO_{2}(\l, N)$}

An exact relation is 
\be
\la{9.1}
 \vevD{:\Omega_{2}:} = -\l\del\Delta F_{N}(\l).
\ee
To prove this relation we simply notice that 
\ba
-\l\del\Delta F_{N} &= \l\del\log\vev{D}_{\rm SYM} = \l\del\log\frac{\int_{\mathfrak{u}(N)} \mc DM\, \exp\bigg(-\frac{8\pi^{2}}{\gym^{2}}\tr M^{2}-\tr L(M)\bigg)}
{\int_{\mathfrak{u}(N)} \mc DM\, \exp\bigg(-\frac{8\pi^{2}}{\gym^{2}}\tr M^{2}\bigg)} \lp
=\frac{8\pi^{2}}{\gym^{2}}[ \vevD{\tr M^{2}}-\vev{\tr M^{2}}_{\rm SYM}] = \vevD{:\tr \Omega_{2}:}.
\ea
From (\ref{9.1}), we immediately obtain the $1/N$ expansion of $\OO_{2}(\l, N)$.
The first two corrections, \ie terms $\sim 1/N$ and $\sim 1/N^{2}$,  are 
\ba
\la{9.3}
\OO_{2}^{(1)} &= \frac{\l^{3}}{16\pi^{2}} (2 \sF_0'+\l  \sF_0''){}^2, \\
\la{9.4}
\OO_{2}^{(2)} &= \frac{\l^{3}}{96\pi^{2}} (-6 \sF_0''+
\l  (-6 \sF_0{}^{(3)}-6 (2 \sF_0'+\l  
\sF_0''){}^2 (2 \sF_0'+\l  (4 \sF_0''
+\l  \sF_0{}^{(3)}))-\l  \sF_0{}^{(4)})),
\ea
and they reproduce $\OO_{2}^{(1)}(\l)$ and $\OO_{2}^{(2)}(\l)$  in (\ref{8.6}).

\subsection*{Results for $\OO_{4}(\l, N)$}

In this case we need derivatives with respect to the couplings $\bm t$ (again here we split the ``kinetic term'' out of $t_{2}$ in $L$) that (for $n>1$) are given by 
\ba
\partial_{t_{2n}}\Delta F_{N} &= \vevD{\tr M^{2n}} = \left(\frac{\gym^{2}}{8\pi^{2}}\right)^{n}\,\vevD{\Omega_{2n}}.
\ea
From the expression of $:\Omega_{4}$ in (\ref{8.2}), 
\ba
\vevD{:\Omega_{4}:} &= \vevD{\Omega_{4}}-2N\,\vevD{:\Omega_{2}:+\frac{N^{2}}{2}}-\vevD{\Omega_{1,1}}+\frac{N(2N^{2}+1)}{4} \lp
= \left(\frac{8\pi^{2}}{\gym^{2}}\right)^{2}\,\partial_{t_{4}}\Delta F_{N}+2N\,\l\del\Delta F_{N}-\vevD{\Omega_{1,1}}-\frac{N(2N^{2}-1)}{4}.
\ea
Using the Toda relation (\ref{C.6}), we can eliminate $\vevD{\Omega_{1,1}}$
\ba
\vevD{:\Omega_{4}:} &= \left(\frac{8\pi^{2}}{\gym^{2}}\right)^{2}\,\partial_{t_{4}}\Delta F_{N}+2N\,\l\del\Delta F_{N}-\frac{N}{2}e^{-\mc D^{2}\Delta F_{N}}-\frac{N(2N^{2}-1)}{4}.
\ea
We now need a Volterra hierachy equation to express $\partial_{t_{4}}\Delta F$. From the results in Appendix \ref{app:volt} with (\ref{B.18}) in (\ref{B.17}), we obtain 
\ba
\vevD{:\Omega_{4}:} &= N^{2}\bigg(\frac{64\pi^{4}}{\l^{2}}\partial_{t_{4}}\sF_{1}+2\l\sF_{0}'\bigg)+\frac{N}{2}(4\l^{2}[\sF_{0}']^{2}+4\l^{3}\sF_{0}'\sF_{0}''+\l^{4}[\sF_{0}'']^{2})+\mc O(N^{0}),
\ea
and using (\ref{2.22})
\be
\OO_{4}(\l, N) = \partial_{t_{4}}\sF_{1}+\frac{\l^{3}}{32\pi^{4}}\sF_{0}'+\frac{1}{N}\frac{\l^{2}}{128\pi^{4}}(4\l^{2}[\sF_{0}']^{2}+4\l^{3}\sF_{0}'\sF_{0}''+\l^{4}[\sF_{0}'']^{2})+\mc O(1/N^{2}).
\ee
Replacing in this expression the expansions (\ref{3.1}) and evaluating $\partial_{t_{4}}\sF_{1}$ by differentiating  (\ref{3.2}) and replacing the couplings in (\ref{2.13}), one reproduces
$\OO_{4}^{(0)}(\l)$ and $\OO_{4}^{(1)}(\l)$ in  (\ref{8.7}).
Of course, the planar term is independently known from (\ref{8.9}), while the $1/N$ correction is a new result
\be
\la{9.10}
\OO_{4}^{(1)}(\l) =\frac{1}{2}\left(\frac{\l}{8\pi^{2}}\right)^{2}\,\l^{2}\,(4\sF_{0}'{}^{2}+4\l\sF_{0}'\sF_{0}''+\l^{2}\sF_{0}''{}^{2}) = \frac{\l}{8\pi^{2}}\OO_{2}^{(1)}(\l).
\ee
The next order, \ie the correction $1/N^{2}$, using (\ref{B.19}) reads
\ba
\la{9.11}
\OO_{4}^{(2)}(\l) &=  
-\frac{1}{12}\left(\frac{\l}{8\pi^{2}}\right)^{2} \l ^2 [96 \l  \sF_0'{}^3+15 
\sF_0''+30 \l ^4 \sF_0''{}^3+6 \l ^5 \sF_0''{}^2 \sF_0{}^{(3)}\lp
+24 \l ^3 \sF_0' 
\sF_0'' (6 \sF_0''+\l  \sF_0{}^{(3)})+24 
\l ^2 \sF_0'{}^2 (9 \sF_0''+\l  
\sF_0{}^{(3)})+\l  (9 \sF_0{}^{(3)}+\l  
\sF_0{}^{(4)})],
\ea
and again can be checked to reproduce the  $1/N^{2}$ contribution $\OO_{4}^{(2)}(\l)$ in (\ref{8.7}).

\subsection*{Results for $\OO_{6}(\l, N)$}

In this case
\ba
\vevD{:\Omega_{6}:} &= \vevD{\Omega_{6}}-3N\,\vevD{\Omega_{4}}-\frac{3}{2}\vevD{\Omega_{2,2}}+
\frac{15}{4}(N^{2}+1)\,\vevD{\Omega_{2}}-3\,\vevD{\Omega_{1,3}}\lp
+\frac{15N}{4}\,\vevD{\Omega_{1,1}}
 -\frac{5}{8}N^{2}(N^{2}+2) \lp
 = \vevD{\Omega_{6}}-3N\,\vevD{\Omega_{4}}-\frac{3}{2}\vevD{\Omega_{2,2}}+
\frac{15}{4}(N^{2}+1)\,\vevD{:\Omega_{2}:}-3\,\vevD{\Omega_{1,3}}\lp
+\frac{15N}{4}\,\vevD{\Omega_{1,1}}
 +\frac{5}{8}N^{2}(2N^{2}+1).
\ea
We can eliminate $\vevD{\Omega_{2,2}}$ by 
\ba
-(\l\del)^{2}\Delta F_{N} &= (\l\del)^{2}\log\vev{D}_{\rm SYM} = (\l\del)^{2}\log\frac{\int_{\mathfrak{u}(N)} \mc DM\, \exp\bigg(-\frac{8\pi^{2}}{\gym^{2}}\tr M^{2}-\tr L(M)\bigg)}
{\int_{\mathfrak{u}(N)} \mc DM\, \exp\bigg(-\frac{8\pi^{2}}{\gym^{2}}\tr M^{2}\bigg)} \lp
= \vevD{\Omega_{2,2}}-\vevD{\Omega_{2}}^{2}+\vevD{\Omega_{2}}-\vev{\Omega_{2,2}}_{\rm SYM}+\vev{\Omega_{2}}_{\rm SYM}^{2}-\vev{\Omega_{2}}_{\rm SYM} \lp
= \vevD{\Omega_{2,2}}-\vevD{\Omega_{2}}^{2}+\vevD{\Omega_{2}}-N^{2}.
\ea
Thus, 
\be
\vevD{\Omega_{2,2}} = -(\l\del)^{2}\Delta F_{N}+[\l\del\Delta F_{N}]^{2}-(N^{2}+1)\l\del\Delta F_{N}+\frac{1}{4}N^{2}(N^{2}+2).
\ee
It remains to eliminate $\vevD{\Omega_{1,3}}$ going back to the Toda recursion. After some work we find in this case
\be
\la{9.15}
\OO_{6}^{(1)}(\l) = \frac{3}{8}\left(\frac{\l}{8\pi^{2}}\right)^{3}\,\l^{3}\,\sF_{0}''\,(4\sF_{0}'+\l\sF_{0}''),
\ee
that  reproduces the  $1/N$ contribution $\OO_{6}^{(1)}(\l)$ in (\ref{8.8}). Also, 
\ba
\OO_{6}^{(2)}(\l) &= \frac{1}{16}\left(\frac{\l}{8\pi^{2}}\right)^{3}\,\l^{2}\,
[-80 \l  \sF_0'{}^3-40 \l ^4 \sF_0''{}^3-6 \l ^5 \sF_0''{}^2 \sF_0{}^{(3)}\lp
-12\,\l ^3 \sF_0' 
\sF_0'' (15 \sF_0''+2\l  \sF_0{}^{(3)})-24
\l ^2 \sF_0'{}^2 (10 \sF_0''+\l  
\sF_0{}^{(3)})-\l  (14 \sF_0{}^{(3)}+\l  \sF_0{}^{(4)})].
\ea

\subsection{Strong coupling expansion}

Let us begin with the planar term (\ref{8.9}). For $n=1$, using (\ref{9.1})
\be
\OO_{2}^{(0)}(\l) = -\frac{\l^{2}}{8\pi^{2}}\del\sF_{0},
\ee
and from (\ref{5.15}),
\be
\la{9.18}
\OO_{2}^{(0)}(\l) = -\frac{\l^{3/2}}{24\pi^{3}}+\frac{\sql}{48\pi}+\cdots.
\ee
For $\OO_{4}^{(0)}$, we use the recursion (\ref{8.11}) 
\ba
\del\OO_{4}^{(0)}(\l) = \frac{1}{8\pi^{2}}\,\l^{3}\frac{d}{d\l}\bigg[\frac{1}{\l^{2}}\OO^{(0)}_{2}(\l)\bigg],
\ea
and obtain 
\be
\OO_{4}^{(0)}(\l) = \frac{\l^{5/2}}{960\pi^{5}}-\frac{\l^{3/2}}{384\pi^{3}}+\frac{7\sql}{1536\pi}+\cdots.
\ee
The general structure, from (\ref{8.11}), turns out to be
\be
\la{9.21}
\OO_{2n}^{(0)}(\l) = 8\,(-1)^{n}\,\left(\frac{\l}{16\pi^{2}}\right)^{n+\frac{1}{2}}\,\frac{n}{4n^{2}-1}\,\bigg[1+\frac{1-4n^{2}}{6}\frac{\pi^{2}}{\l}+\frac{7(1-4n^{2})^{2}}{360}\frac{\pi^{4}}{\l^{2}}+\cdots\bigg].
\ee
The next corrections can be computed by using the exact expressions (\ref{9.3}, \ref{9.4}) and (\ref{9.10}, \ref{9.11}). For $\OO_{2}(\l)$ we have 
\ba
\la{9.22}
\OO_{2}^{(1)}(\l) &= \frac{\l^{2}}{64\pi^{4}}-\frac{\l}{192\pi^{2}}-\frac{1}{720}+\cdots, \\
\OO_{2}^{(2)}(\l) &= -\frac{\l^{5/2}}{256\pi^{5}}+\frac{\l^{3/2}}{512\pi^{3}}+\frac{49\sql}{92160\pi}+\cdots, \notag
\ea
Using (\ref{9.1}), (\ref{5.17}), and (\ref{5.20}), we can resum the leading terms at large tension as 
\be
\OO_{2}(\l; N) \stackrel{\rm LT}{=}  N^{3}f_{2}\left(\frac{16\pi^{2}N^{2}}{\l}\right), \qquad f_{2}(x) = \frac{8\pi^{2}}{3x^{3}}+\frac{4}{x^{2}}-\frac{8}{3x^{3}}(1+x)^{3/2}.
\ee
As a check, expanding at large $N$ we find 
\be
\OO_{2}(\l; N) \stackrel{\rm LT}{=} -\frac{\l ^{3/2}}{24 \pi ^3}+\frac{\l ^2}{64 \pi ^4 
N}-\frac{\l ^{5/2}}{256 \pi ^5 N^2}+\frac{\l ^3}{1536 \pi 
^4 N^3}-\frac{\l ^{7/2}}{16384 \pi ^7 N^4}+\cdots, 
\ee
that reproduces the leading terms at large $\l$ in (\ref{9.18})  and (\ref{9.22}).

\noindent
For $\OO_{4}(\l)$ we find
\ba
\la{9.25}
\OO_{4}^{(1)}(\l) &= \frac{\l^{3}}{512\pi^{6}}-\frac{\l^{2}}{1536\pi^{4}}-\frac{\l}{5760\pi^{2}}-\frac{13}{40320}+\cdots, \\
\OO_{4}^{(2)}(\l) &= -\frac{3\l^{7/2}}{2048\pi^{7}}+\frac{7\l^{5/2}}{4096\pi^{5}}-\frac{49\l^{3/2}}{147456\pi^{3}}-\frac{1037\sql}{30965760\pi}+\cdots\, .\notag
\ea
For $\OO_{6}(\l)$ we find
\ba
\OO_{6}^{(1)}(\l) &= -\frac{7 \l ^4}{49152 \pi ^8}+\frac{13 \l ^3}{49152 \pi 
^6}-\frac{13 \l ^2}{61440 \pi ^4}-\frac{43 \l }{1290240 \pi 
^2}-\frac{289}{1290240}+\cdots, \\
\OO_{6}^{(2)}(\l) &= -\frac{3 \l ^{9/2}}{65536 \pi ^9}-\frac{167 \l
^{7/2}}{393216 \pi ^7}+\frac{20851 \l ^{5/2}}{23592960 \pi 
^5}-\frac{563657 \l ^{3/2}}{990904320 \pi ^3}-\frac{5922863 
\sql}{13212057600 \pi }\cdots\,.
\ea
A cleaner way of presenting these expansions is in terms of the scaled coupling $\hat{\l} = \frac{\l}{4\pi^{2}}$
\ba
\OO_{2}^{(1)}(\l) &= \frac{1}{4}\,\hat{\l}^{2}\,\left(1-\frac{1}{12\,\hat\l}-\frac{1}{180\,{\hat\l}^{2}}+\cdots\right),\\
\la{9.29}
\OO_{2}^{(2)}(\l) &= -\frac{1}{8}\,\hat{\l}^{5/2}\,\left(1-\frac{1}{8\,\hat\l}-\frac{49}{5760\,\hat{\l}^{2}}+\cdots\right),\\
\OO_{4}^{(1)}(\l) &= \frac{1}{8}\,\hat{\l}^{3}\,\left(1-\frac{1}{12\,\hat\l}-\frac{1}{180\,{\hat\l}^{2}}-\frac{13}{5040\,\hat{\l}^{3}}+\cdots\right),\\
\la{9.31}
\OO_{4}^{(2)}(\l) &= -\frac{3}{16}\,\hat{\l}^{7/2}\,\left(1-\frac{7}{24\,\hat\l}+\frac{49}{3456\,\hat{\l}^{2}}+\frac{1037}{2903040\,\hat{\l}^{3}}+\cdots\right), \\
\OO_{6}^{(1)}(\l) &= -\frac{7}{192}\,\hat{\l}^{4}\,\left(1-\frac{13}{28\,\hat\l}+\frac{13}{140\,\hat{\l}^{2}}+\frac{43}{11760\,\hat{\l}^{3}}
+\frac{289}{47040\,\hat{\l}^{4}}+\cdots\right), \\
\OO_{6}^{(2)}(\l) &= -\frac{3}{128}\,\hat{\l}^{9/2}\,\left(
1+\frac{167}{72\,\hat{\l}}-\frac{20851}{17280\,\hat{\l}^{2}}+\frac{563657}{2903040\,\hat{\l}^{3}}+\frac{5922863}{154828800\,\hat{\l}^{4}}+\cdots
\right).
\ea

\section{One-point functions in the $SU(N)$ model}
\la{sec:one-sun}

In the $SU(N)$ case, the explicit expansions written in (\ref{8.6}) , (\ref{8.7}) , (\ref{8.8}) can be recomputed with the traceless constraint
and one finds  \footnote{We use a tilde to denote $SU(N)$ quantities.}
\be
\tilde\OO_{n}^{(k)}(\l) = \OO_{n}^{(k)}(\l), \qquad k=0,1, \ n=2,4,6,\dots,
\ee
showing that it gives deviations with (relative) size $1/N^{2}$. The one-point functions at this order are 
\ba
\tilde\OO_{2}^{(2)}(\l) &= \frac{\l}{8\pi^{2}}\,\bigg(
\frac{\l }{32}-\frac{5 \l ^2}{1536}+\frac{5 \l ^3}{24576}+\frac{\l 
^4}{589824}-\frac{191 \l ^5}{56623104}+\cdots\bigg), \\
\tilde\OO_{4}^{(2)}(\l) &= \left(\frac{\l}{8\pi^{2}}\right)^{2}\,\bigg(
-\frac{7 \l ^2}{3072}+\frac{409 \l ^4}{11796480}-\frac{71 \l 
^5}{8847360}+\frac{21481 \l ^6}{15854469120}-\frac{27143 \l 
^7}{135895449600}
+\cdots\bigg), \\
\tilde\OO_{6}^{(2)}(\l) &= \left(\frac{\l}{8\pi^{2}}\right)^{3}\,\bigg(
-\frac{17 \l ^3}{98304}+\frac{23 \l ^4}{491520}-\frac{4583 
\l ^5}{528482304}+\frac{85937 \l^6}{63417876480}-\frac{117013 \l ^7}{608811614208}
+\cdots\bigg), \\
\tilde\OO_{8}^{(2)}(\l) &= \left(\frac{\l}{8\pi^{2}}\right)^{4}\,\bigg(
\frac{59 \l ^4}{4718592}-\frac{433 \l ^5}{123863040}+\frac{125071 \l 
^6}{190253629440}-\frac{197261 \l ^7}{1902536294400}
+\cdots\bigg), \\
\ea
and  they differ from the corresponding $U(N)$ expressions. Nevertheless, the case of $\tilde\OO_{2}$ is still captured by (\ref{9.1}). From that relation 
and (\ref{6.2}) one obtains
\be
\la{10.7}
\tilde \OO_{2}^{(2)}(\l)- \OO_{2}^{(2)}(\l) = \frac{\l^{2}}{16\pi^{2}}[\l\,(\l\,\sF_{0})'']'.
\ee
In the case of $\tilde\OO_{4}$ we found \footnote{We used the fact that, in a general model, the shift is linear in $t_{2n}$ and therefore
in $\sF_{0}$.} 
\be
\tilde \OO_{4}^{(2)}(\l)- \OO_{4}^{(2)}(\l) = \left(\frac{\l}{8\pi^{2}}\right)^{2}\,\bigg(3\l^{2}\sF_{0}''+\frac{1}{2}\l^{3}\sF_{0}'''\bigg).
\ee
Starting with $\OO_{6}$ the structure changes and we have
\ba
\l^{2}\del\bigg[\frac{1}{\l}(\tilde\OO_{6}^{(2)}-\OO_{6}^{(2)})\bigg] &= \left(\frac{\l}{8\pi^{2}}\right)^{3}\bigg(\frac{15}{4}\l^{3}\,\del^{3}\sF_{0}+\frac{3}{8}\l^{4}\del^{4}\sF_{0}\bigg), \\
(\l^{2}\del)^{2}\bigg[\frac{1}{\l^{2}}(\tilde\OO_{8}^{(2)}-\OO_{8}^{(2)})\bigg] &= \left(\frac{\l}{8\pi^{2}}\right)^{4}\bigg(\frac{7}{2}\l^{4}\,\del^{4}\sF_{0}+\frac{1}{4}\l^{5}\del^{5}\sF_{0}\bigg), \\
(\l^{2}\del)^{3}\bigg[\frac{1}{\l^{3}}(\tilde\OO_{10}^{(2)}-\OO_{10}^{(2)})\bigg] &= \left(\frac{\l}{8\pi^{2}}\right)^{5}\bigg(\frac{45}{16}\l^{5}\,\del^{5}\sF_{0}+\frac{5}{32}\l^{6}\del^{6}\sF_{0}\bigg).
\ea
suggesting the general form, valid for $n\ge 2$
\be
\la{10.12}
(\l^{2}\del)^{n-2}\bigg[\frac{1}{\l^{n-2}}(\tilde\OO_{2n}^{(2)}-\OO_{2n}^{(2)})\bigg] = \frac{1}{2^{n-1}}\left(\frac{\l}{8\pi^{2}}\right)^{n}\bigg[n(2n-1)\l^{n}\,\del^{n}\sF_{0}+\frac{n}{2}\l^{n+1}\del^{n+1}\sF_{0}\bigg].
\ee
that we checked for higher values of $n$.

\subsection{Deviations in the strong coupling expansion}

At strong coupling (we use the same notation as in (\ref{9.29}) and (\ref{9.31}) for comparison)
\ba
\tilde \OO_{2}^{(2)}(\l)- \OO_{2}^{(2)}(\l) &= \frac{\l^{3/2}}{64\pi^{3}}+\frac{\sql}{384\pi}+\cdots =  -\frac{1}{8}\,\hat{\l}^{5/2}\,\left(0-\frac{1}{\hat\l}-\frac{1}{24\,\hat{\l}^{2}}+\cdots\right),\\ \\
\tilde \OO_{4}^{(2)}(\l)- \OO_{4}^{(2)}(\l) &=-\frac{3\l^{5/2}}{512\pi^{5}}+\frac{7\l^{3/2}}{1024\pi^{3}}-\frac{35\sql}{12288\pi}+\cdots = 
-\frac{3}{16}\,\hat{\l}^{7/2}\,\left(0+\frac{1}{\hat\l}-\frac{7}{24\,\hat{\l}^{2}}+\cdots\right),
\ea
and, in both cases, the correction is subleading compared with the leading term in (\ref{9.29}) and (\ref{9.31}). The pattern continues for higher one-point functions, using (\ref{10.12}) and in all cases
the correction is still subleading.

\section{Conclusions and open issues}
\la{sec:conclusions}

We have considered the fluxless  D3-D5 system and the dual dCFT consisting in four dimensional $U(N)$ $\mathcal N=4$ SYM 
in the presence of a codimension-one interface hosting a 3d $\mathcal N=4$ theory. For the free energy and the simplest one-point functions
of BPS scalars, localization reduces their computation to the analysis of a hermitian one-matrix model with a non-polynomial (subleading) 
single-trace potential. Such a matrix model is known to be closely related to integrable hierarchies, \ie the Toda lattice and its Volterra reduction.
We have shown how to exploit this rich structure in order to  analytically control the $1/N$ expansion. In particular, it is possible to 
access the strong coupling regime and determine some features emerging in the large-tension limit where the 't Hooft coupling is taken large
order by order in $1/N$. 

\medskip
While the analysis of the free energy is complete in the sense that can be pushed without effort to higher orders in $1/N$, the situation is 
different for the one-point functions. Despite the results presented at low operator dimension $\Delta$, it is clear that mixing remains a major complication. In particular, 
refined tools are needed in order to derive results depending in closed form on $\Delta$ that could be useful, for instance,  to explore large charge limits.
Another puzzling technical aspect are the complications that are met while extending the analysis from $U(N)$ to $SU(N)$  symmetry on the gauge side.
Their general solution calls for a better physical understanding of the interaction between the extra $U(1)$ degrees of freedom and the 3d theory on the defect.

\medskip
More conceptually, a major issue is to understand the strong coupling expansions of the free energy and one-point functions from the 
perspective of the dual string theory. Generally speaking, since $1/N$ corrections are associated with higher genus string calculations, one 
can hope at least to understand the structure of the strong coupling expansions (leading powers at large $\l$) and their large-tension resummations in the spirit of \cite{Giombi:2020mhz,Beccaria:2020ykg}.
From this point of view, it would also be interesting to further explore the nature (stability) of the non-perturbative corrections expected to 
affect the free energy at strong coupling as well as their resurgence structure, see for instance \cite{Beccaria:2021ism,Beccaria:2022ypy,Beccaria:2022kxy}.

\medskip
At a more practical level, natural extensions of this work concern the generalization to the case of D3-D5 system with flux in order to explore
nicer AdS/CFT limits involving  large fluxes $k\gg 1$. Also, it would be important to 
extend the analysis to defect theories with $\mc N=2$ superconformal symmetry, where the same localization approach as in \cite{Wang:2020seq,Komatsu:2020sup} is expected to 
work.

\section*{Acknowledgments}
We thank  M. Dedushenko, G. Linardopoulos, 
G. Akemann, P. Vivo, and A. Moro for useful discussions. We  acknowledge financial support from the INFN grant GSS (Gauge Theories, Strings and Supergravity). 

\appendix

\section{On the normalization of $Z^{\rm D5}$}
\la{app:norm}

The partition function (\ref{2.7}) has a non-trivial normalization that depends on $\gym$ even in the case $k=0$, \cf (\ref{2.9}). To understand this factor, let us recall that 
(\ref{2.7}) is obtained, in general case with flux,  by integrating out the $U(N+k)$ eigenvalues in the expression in Eq.~(2.28) of \cite{Komatsu:2020sup} (here $g\equiv \gym$)
\ba
Z^{\rm D5}_{N,k} &= \frac{1}{N!(N+k)!}\int\prod_{n=1}^{N}da_{n}\prod_{n=1}^{N+k}db_{n}\, \frac{\Delta(a)\,\widetilde\Delta(a)\,e^{-\frac{g^{2}}{4}\sum_{n=1}^{N}a_{n}^{2}}\ 
\Delta(b)\,\widetilde\Delta(b)\,e^{-\frac{g^{2}}{4}\sum_{n=1}^{N+k}b_{n}^{2}}}{\prod_{n=1}^{N}\prod_{m=1}^{N+k}2\cosh \pi(a_{n}-b_{m})},
\ea
This expression has  a factored form, corresponding to gluing $U(N)$ and $U(N+M)$ SYMs with Neumann boundary conditions and coupled together by the interface $U(N)\times U(N+k)$
bifundamental hypermultiplet. In the above, $\Delta(x) = \prod_{n<m}(x_{n}-x_{m})$ is the Vandermonde determinant, while 
$\widetilde\Delta(x) = \prod_{i<j}2\sinh \pi(x_{i}-x_{j})$
is the one-loop determinant for the 3d $\N=4$ vector multiplets. 

It may be instructive to consider the case $N=1$ and $k=0$. In this case, we have simply
\be
Z^{\rm D5}_{1,0} = \int_{-\infty}^{\infty}da\,\int_{-\infty}^{\infty}db\,\frac{e^{-\frac{g^{2}}{4}(a^{2}+b^{2})}}{2\cosh \pi(a-b)} = 
\left(\frac{\pi}{2g^{2}}\right)^{1/2}\,\int_{-\infty}^{\infty}da\,  \frac{e^{-\frac{g^{2}}{8}\,a^{2}}}{\cosh \pi a},
\ee
where we shifted $b\to b+a$ and integrated over $b$.
Using now
\be
\frac{1}{\cosh \pi a} = \int_{-\infty}^{\infty}dt\, \frac{e^{2\pi ita}}{\cosh \pi t},
\ee
we get
\be
Z^{\rm D5}_{1,0} = \left(\frac{\pi}{2g^{2}}\right)^{1/2}\,\int_{-\infty}^{\infty}da \int_{-\infty}^{\infty}dt\,  \frac{e^{-\frac{g^{2}}{8}\,a^{2}+2\pi i a t}}{\cosh \pi t} = \frac{4\pi}{g^{2}}\int_{-\infty}^{\infty}dt\, \frac{e^{-\frac{8\pi^{2}}{g^{2}}
t^{2}}}{2\cosh \pi t}, 
\ee
which is (\ref{2.9}), taking into account the explicit factors of 2 and $\pi$ in (\ref{2.8}). This very simple special case illustrates why the effect of the defect is not just 
the insertion of the single trace $\log\cosh$ potential, but also the additional pre-factor $(1/\gym^{2})^{N^{2}}$ in the partition function normalization in (\ref{2.18}).

\section{Higher order flows in the Volterra hierarchy}
\la{app:volt}

\subsection{Structure of the flows}

In the notation of Section \ref{sec:hier} we can introduce the following ratios of partition functions
\be
B_{n}(\bm t) = \frac{Z_{n+1}(\bm t)Z_{n-1}(\bm t)}{Z_{n}^{2}(\bm t)} = e^{-\mc D_{+}f_{n}(\bm t)}.
\ee
From the results of  \cite{Alvarez-Gaume:1990asn,adler1995}  we have the equations
\be
\la{B.2}
\partial_{t_{2k}} B_{n}(\bm t) = -B_{n}(\bm t) [V^{(2k)}_{n+1}(\bm t)-V^{(2k)}_{n-1}(\bm t)],
\ee
where (omitting the argument $\bm t$) $V^{(2k)}_{n}$ may be derived by the Lax formalism
\be
V_{n}^{(2k)} = \sqrt{B_{n}}(L^{2k-1})_{n,n+1}, \qquad 
L = \begin{pmatrix} 
0 & \sqrt{B_{1}} & 0 & 0 & \cdots \\
\sqrt{B_{1}} & 0 & \sqrt{B_{2}} & 0 & \cdots \\
0 & \sqrt{B_{2}} & 0 & \sqrt{B_{3}}  & \cdots \\
&& \cdots
\end{pmatrix}\, .
\ee
Explicitly, the first cases are
\ba
V^{(2)}_{n} &= B_{n}, \\
V^{(4)}_{n} &= V_{n}^{(2)}(V_{n-1}^{(2)}+V_{n}^{(2)}+V_{n+1}^{(2)}), \\
V^{(6)}_{n} &= V_{n}^{(2)}(V_{n-1}^{(2)}V_{n+1}^{(2)}+V_{n-1}^{(4)}+V_{n}^{(4)}+V_{n+1}^{(4)}), \\
V^{(8)}_{n} &= B_{n}\,(B_{n-1}^3+2 B_{n-2} B_{n-1}^2+3 B_n B_{n-1}^2+B_{n+1} B_{n-1}^2+B_{n-2}^2
   B_{n-1}+3 B_n^2 B_{n-1}\lp
   +B_{n+1}^2 B_{n-1}+B_{n-3} B_{n-2} B_{n-1}+2
   B_{n-2} B_n B_{n-1}+B_{n-2} B_{n+1} B_{n-1}+4 B_n B_{n+1}
   B_{n-1}\lp
   +B_{n+1} B_{n+2} B_{n-1}+B_n^3+B_{n+1}^3+3 B_n
   B_{n+1}^2+B_{n+1} B_{n+2}^2+3 B_n^2 B_{n+1}+2 B_{n+1}^2 B_{n+2}\lp
   +2 B_n
   B_{n+1} B_{n+2}+B_{n+1} B_{n+2} B_{n+3}),
\ea
and so on. These are the generalizations of (\ref{4.10}) which is simply the $k=1$ case of (\ref{B.2}). Indeed, in this case it reads, \cf (\ref{4.7}),
\be
\partial_{t_{2}}\mc D_{+}f_{n} = e^{-\mc D_{+}f_{n+1}}- e^{-\mc D_{+}f_{n-1}} = r_{n+2}-r_{n} = \mc D_{+}(r_{n+1}+r_{n}).
\ee
This gives
\be
\partial_{t_{2}}f_{n} = r_{n+1}+r_{n}+\text{const}.
\ee
The constant is shown to be zero taking $n=1$ and this gives back (\ref{4.10}).
In our problem, 
  \be
  \la{B.10}
  Z_{N}(\bm t) = 2^{-N^{2}/2}G(N+1) \vev{D}_{\rm SYM}, 
  \ee
  where we do not write the dependence on $\gym$ and as usual $t_{2}$ is from the defect term $L$.
  Since $F_{N} = -\log \vev{D}_{\rm SYM}$, we have 
\be
\la{B.11}
B_{N}(\bm t) = \frac{N}{2}e^{-\mc D^{2}\Delta F_{N}(\bm t)},
\ee
and taking into account the coupling dependent factors attached to the couplings $t_{2n}$, the flows in (\ref{B.2}) should be written
\be
\left(  \frac{8\pi^{2}}{\gym^{2}}\right)^{k}\partial_{t_{2k}} B_{N}(\bm t) = -B_{N}(\bm t) [V^{(2k)}_{N+1}(\bm t)-V^{(2k)}_{N-1}(\bm t)].
\ee

\subsection{The $k=2$ flow and its consequences}

We have discussed the $k=1$ equation in the main text, here let us write the $k=2$ case. It is 
\be
\left(  \frac{8\pi^{2}}{\gym^{2}}\right)^{2}\partial_{t_{4}} B_{N}= -B_{N}[B_{N+1}(B_{N}+B_{N+1}+B_{N+2})-B_{N-1}(B_{N-2}+B_{N-1}+B_{N})],
\ee
that is 
\be
\left(  \frac{8\pi^{2}}{\gym^{2}}\right)^{2}\partial_{t_{4}} \mc D^{2}\Delta F_{N}= B_{N+1}(B_{N}+B_{N+1}+B_{N+2})-B_{N-1}(B_{N-2}+B_{N-1}+B_{N}),
\ee
Replacing here (\ref{B.11}) and the expansion (\ref{2.17}) with rules like (\ref{5.6}) or (\ref{5.7}), we obtain 
\ba
\partial_{t_{4}}\sF_{0}(\l) &= \frac{\l^{2}}{128\pi^{4}}, \\
\partial_{t_{4}}\sF_{1}'(\l) &= -\frac{3\l^{2}}{64\pi^{4}}(2\sF_{0}'+\l\sF_{0}''),\\
\la{B.17}
\partial_{t_{4}}\sF_{2}(\l) &= c_{1}\l+c_{2}\l^{2}+\frac{3\l^{4}}{128\pi^{4}}(2\sF_{0}'+\l\sF_{0}'')^{2},
\ea
where we used systematically the differential relations for $\sF_{n}$, as in (\ref{5.8}) etc. In the D3-D5 system, comparison with weak coupling gives
\be
\la{B.18}
c_{1}=0, \qquad c_{2}=\frac{1}{256\pi^{4}}.
\ee
The next relation is substantially more involved.  Zero modes are absent in the D3-D5 system and 
one finds
\ba
\la{B.19}
\partial_{t_{4}}\sF_{3}(\l) &= -\frac{\l ^3}{256 \pi ^4} (4 \sF_0'+64 \l ^2 \sF_0'{}^3+11 \l  \sF_0''+168 \l ^3 \sF_0'{}^2 
\sF_0''+120 \l ^4 \sF_0' \sF_0''{}^2+26 
\l ^5 \sF_0''{}^3\lp
+7 \l ^2 \sF_0{}^{(3)}+24 
\l ^4 \sF_0'{}^2 \sF_0{}^{(3)}+24 \l ^5 
\sF_0' \sF_0'' \sF_0{}^{(3)}+6 \l ^6 
\sF_0''{}^2 \sF_0{}^{(3)}+\l ^3 
\sF_0{}^{(4)}).
\ea

\section{Toda $t_{1}$ flow and $1/N$ expansion of $\vevD{\Omega_{1, \dots, 1}}$}
\la{app:t1}

Let us begin by writing the Toda equation (\ref{4.9}) in the form 
\ba
 \mc D^{2}F_{N}(\bm t) &= -\log\bigg(-\frac{\partial^{2}}{\partial t_{1}^{2}}F_{N}(\bm t)\bigg).
 \ea
Since
\be
\Delta F_{N} = F_{N}-F_{N}^{\rm Gaussian} = F_{N}+\log\bigg[(2\pi)^{-\frac{N}{2}}\,G(N+1)\,\left(\frac{16\pi^{2}}{\gym^{2}}\right)^{-\frac{N^{2}}{2}}\bigg]
\ee
we obtain 
\be
 \mc D^{2}\Delta F_{N}(\bm t) -\log N+\log\left(\frac{16\pi^{2}}{\gym^{2}}\right)= -\log\bigg(-\frac{\partial^{2}}{\partial t_{1}^{2}}\Delta  F_{N}(\bm t)\bigg),
 \ee
 that we write as  \footnote{The l.h.s. of (\ref{C.4}) should be expressed in terms of $\gym$ before applying $\mc D^{2}$.}
\be
\la{C.4}
 \mc D^{2}\Delta F_{N}(\bm t) = -\log\bigg(-\frac{16\pi^{2}}{\l}\frac{\partial^{2}}{\partial t_{1}^{2}}\Delta  F_{N}(\bm t)\bigg),
\ee
Differentiating  (\ref{C.4}), we obtain  formulas for $\vevD{\Omega_{1,\dots, 1}}$ in terms of $\sF_{0}(\l)$. Let us see the first two examples.

\subsection{$\vevD{\Omega_{1,1}}$}

From
\be
\la{C.5}
\frac{\partial^{2}}{\partial t_{1}^{2}}\Delta  F_{N}(\bm t)\bigg|_{t_{1}=0} =-\frac{\l}{8\pi^{2}N} \vevD{\Omega_{1,1}},
\ee
we have 
\be
\la{C.6}
\vevD{\Omega_{1,1}} = \frac{N}{2}\,e^{-\mc D^{2}\Delta F_{N}(\bm t)}.
 \ee
 If we parametrize
 \be
 \la{C.7}
 \vevD{\Omega_{1,1}} = \frac{N}{2}+\omega_{1,1}^{(1)}(\l)+\frac{1}{N}\omega_{1,1}^{(2)}(\l)+\cdots, 
 \ee
 and use (\ref{5.6}),
 we obtain differential relations. The first of them gives
 \be
 \la{C.8}
 \omega_{1,1}^{(1)}(\l) = -\l\sF_{0}'-\frac{\l^{2}}{2}\sF_{0}'' = -\frac{\l}{2}\,(\l \sF_{0})''.
 \ee
 and integrating by parts (\ref{3.8}) we obtain 
 \be
 \omega_{1,1}^{(1)}(\l) = -\oint\frac{dx}{2\pi i}\frac{x}{(1-x^{2})^{2}}\log\cosh\bigg[\frac{\sql}{4}(x+x^{-1})\bigg].
 \ee
 For the coefficient $\omega_{1,1}^{(2)}(\l)$ we find 
 \ba
 \omega_{1,1}^{(2)}(\l) &= [\omega_{1,1}^{(1)}(\l)]^{2}-\frac{\l^{2}}{2}\sF_{1}'' = \bigg(-\l\sF_{0}'-\frac{\l^{2}}{2}\sF_{0}''\bigg)^{2}-\frac{\l^{2}}{2}\del\bigg[-\frac{1}{2}\l[(\l\sF_{0})'']^{2}\bigg],
 \ea
 and comparing with (\ref{C.8}) we get the simple relation
 \be
 \la{C.11}
 \omega_{1,1}^{(2)}=\l\partial_{\l}[(\omega_{1,1}^{(1)})^{2}].
 \ee

\subsection{$\vevD{\Omega_{1,1,1,1}}$}

The next example is 
\be
\frac{\partial^{4}}{\partial t_{1}^{4}}\Delta F_{N}(\bm t)\bigg|_{t_{1}=0} = -\left(\frac{\l}{8\pi^{2}N}\right)^{2}[\vevD{\Omega_{1,1,1,1}}-3\vevD{\Omega_{1,1}}^{2}].
\ee
From (\ref{C.4}),
\ba
\frac{\partial^{4}}{\partial t_{1}^{4}}\Delta F_{N}(\bm t) \bigg|_{t_{1}=0} &= -\frac{\l}{16\pi^{2}}\frac{\partial^{2}}{\partial t_{1}^{2}}e^{-\mc D^{2}\Delta F_{N}(\bm t)}
\bigg|_{t_{1}=0} = \frac{\l}{16\pi^{2}}\frac{\partial^{2}}{\partial t_{1}^{2}}\mc D^{2}\Delta F_{N}(\bm t)\, e^{-\mc D^{2}\Delta F_{N}}\bigg|_{t_{1}=0} \lp
= -\frac{\l}{(16\pi^{2})^{2}}e^{-\mc D^{2}\Delta F_{N}}\mc D^{2}\bigg[\l e^{-\mc D^{2}\Delta F_{N}(\bm t)}\bigg].
\ea
This gives
\be
\vevD{\Omega_{1,1,1,1}}-3\vevD{\Omega_{1,1}}^{2} = \frac{N^{2}}{4\l}e^{-\mc D^{2}\Delta F_{N}}\mc D^{2}\bigg[\l e^{-\mc D^{2}\Delta F_{N}(\bm t)}\bigg].
\ee
Replacing the expansion of $F_{N}$ (\ref{2.17}), we obtain 
\be
\la{C.15}
\vevD{\Omega_{1,1,1,1}} =  \frac{3N^{2}}{4}+3N\omega_{1,1}^{(1)}(\l)+3\del[\l(\omega_{1,1}^{(1)})^{2}]+\cdots.
\ee

\subsection{General $\vevD{\Omega_{1,\dots 1}}$ at order $1/N^{2}$}

Inspection of the next cases provides the following simple generalization to the case with $2n$ indices ``1''
\be
\vevD{(\tr A)^{2n}} =  \frac{(2n-1)!!}{2^{n}}\, N^{n}\bigg[1+\frac{2n}{N}\omega_{1,1}^{(1)}(\l)
+\frac{2n}{N^{2}}\frac{1}{\l^{n-2}}\del[\l^{n-1}(\omega_{1,1}^{(1)})^{2}]+\cdots\bigg].
\ee

\section{Compact expressions for the free energy in the $U(N)$ model}
\la{app:ZZZ}
The explicit expression of the free energy expansion coefficients $\sF_{n}(\l)$ in (\ref{2.17}) have been discussed in Section \ref{sec:free-exp}.
Here, we provide the simplified expresssions that are obtained in terms of the quantity $Z(\l)$ defined in (\ref{5.12}) and up to $\sF_{6}(\l)$.
\ba
\sF_{1}'(\l) &= -\frac{1}{2\l}\,Z^{2}, \\
\sF_{2}(\l) &= -\frac{Z}{12}+\frac{Z^3}{6}+\frac{1}{12} \l  Z' , \\
\sF_{3}(\l) &= -\frac{1}{24} Z^2+\frac{Z^4}{24}+\frac{1}{12} \l  Z Z'-\frac{1}{6} \l 
 Z^3 Z'-\frac{1}{24} \l ^2 Z'^2-\frac{1}{12} \l ^2 Z Z'' , \\
\sF_{4}(\l) &= \frac{Z}{120}-\frac{Z^3}{36}+\frac{Z^5}{60}-\frac{1}{120} \l  
Z'+\frac{1}{12} \l  Z^2 Z'-\frac{1}{12} \l  Z^4 Z'-\frac{1}{12} \l ^2 
Z Z'^2+\frac{1}{6} \l ^2 Z^3 Z'^2\lp
+\frac{1}{36} \l ^3 
Z'^3+\frac{1}{240} \l ^2 Z''-\frac{1}{24} \l ^2 Z^2 Z''+\frac{1}{24} 
\l ^2 Z^4 Z''+\frac{1}{6} \l ^3 Z Z' Z''-\frac{1}{720} \l ^3 Z^{(3)}\lp
+\frac{1}{24} \l ^3 Z^2 Z^{(3)}+\frac{1}{288} \l ^4 Z^{(4)}, \\
\sF_{5}(\l) &= \frac{Z^2}{80}-\frac{Z^4}{48}+\frac{Z^6}{120}-\frac{1}{40} \l  Z 
Z'+\frac{1}{12} \l  Z^3 Z'-\frac{1}{20} \l  Z^5 Z'+\frac{1}{80} \l ^2 
Z'^2-\frac{1}{8} \l ^2 Z^2 Z'^2 \lp
+\frac{1}{8} \l ^2 Z^4 
Z'^2+\frac{1}{12} \l ^3 Z Z'^3-\frac{1}{6} \l ^3 Z^3 
Z'^3-\frac{1}{48} \l ^4 Z'^4+\frac{1}{80} \l ^2 Z Z''-\frac{1}{24} \l 
^2 Z^3 Z'' \lp
+\frac{1}{40} \l ^2 Z^5 Z''-\frac{1}{80} \l ^3 Z' 
Z''+\frac{1}{8} \l ^3 Z^2 Z' Z''-\frac{1}{8} \l ^3 Z^4 Z' 
Z''-\frac{1}{4} \l ^4 Z Z'^2 Z''+\frac{1}{320} \l ^4 
Z''^2 \lp
-\frac{1}{12} \l ^4 Z^2 Z''^2-\frac{1}{240} \l ^3 Z 
Z^{(3)}+\frac{1}{72} \l ^3 Z^3 Z^{(3)}-\frac{1}{120} \l ^3 Z^5 
Z^{(3)}+\frac{1}{240} \l ^4 Z' Z^{(3)}\lp
-\frac{1}{8} \l ^4 Z^2 Z' 
Z^{(3)}-\frac{29 \l ^5 Z'' Z^{(3)}}{1440}-\frac{1}{480} \l ^4 Z 
Z^{(4)}-\frac{1}{72} \l ^4 Z^3 Z^{(4)}-\frac{1}{96} \l ^5 Z' Z^{(4)}-
\frac{1}{288} \l ^5 Z Z^{(5)}, \\
\sF_{6}(\l) &= -\frac{Z}{252}+\frac{Z^3}{60}-\frac{Z^5}{60}+\frac{Z^7}{210}+\frac{1}{
252} \l  Z'-\frac{1}{20} \l  Z^2 Z'+\frac{1}{12} \l  Z^4 
Z'-\frac{1}{30} \l  Z^6 Z'  \lp
+\frac{1}{20} \l ^2 Z Z'^2-\frac{1}{6} \l 
^2 Z^3 Z'^2+\frac{1}{10} \l ^2 Z^5 Z'^2-\frac{1}{60} \l ^3 
Z'^3+\frac{1}{6} \l ^3 Z^2 Z'^3-\frac{1}{6} \l ^3 Z^4 
Z'^3 \lp
-\frac{1}{12} \l ^4 Z Z'^4+\frac{1}{6} \l ^4 Z^3 
Z'^4+\frac{1}{60} \l ^5 Z'^5-\frac{1}{504} \l ^2 Z''+\frac{1}{40} \l 
^2 Z^2 Z''-\frac{1}{24} \l ^2 Z^4 Z'' \lp
+\frac{1}{60} \l ^2 Z^6 
Z''-\frac{1}{20} \l ^3 Z Z' Z''+\frac{1}{6} \l ^3 Z^3 Z' 
Z''-\frac{1}{10} \l ^3 Z^5 Z' Z''+\frac{1}{40} \l ^4 Z'^2 Z''-\frac{1}{4} \l ^4 Z^2 Z'^2 Z'' \lp
+\frac{1}{4} \l ^4 Z^4 Z'^2 
Z''+\frac{1}{3} \l ^5 Z Z'^3 Z''+\frac{1}{80} \l ^4 Z 
Z''^2-\frac{1}{24} \l ^4 Z^3 Z''^2+\frac{1}{40} \l ^4 Z^5 
Z''^2-\frac{1}{80} \l ^5 Z' Z''^2  \lp
+\frac{1}{3} \l ^5 Z^2 Z' 
Z''^2+\frac{7}{360} \l ^6 Z''^3+\frac{\l ^3 
Z^{(3)}}{1512}-\frac{1}{120} \l ^3 Z^2 Z^{(3)}+\frac{1}{72} \l ^3 Z^4 
Z^{(3)}-\frac{1}{180} \l ^3 Z^6 Z^{(3)} \lp
+\frac{1}{60} \l ^4 Z Z' 
Z^{(3)}-\frac{1}{18} \l ^4 Z^3 Z' Z^{(3)}+\frac{1}{30} \l ^4 Z^5 Z' 
Z^{(3)}-\frac{1}{120} \l ^5 Z'^2 Z^{(3)}+\frac{1}{4} \l ^5 Z^2 Z'^2 
Z^{(3)} \lp
+\frac{7}{720} \l ^5 Z Z'' Z^{(3)}+\frac{7}{72} \l ^5 Z^3 Z'' 
Z^{(3)}+\frac{29}{360} \l ^6 Z' Z'' Z^{(3)}+\frac{29 \l ^6 Z 
Z^{(3)}{}^2}{1440}-\frac{\l ^4 Z^{(4)}}{6048}  \lp
+\frac{1}{480} \l ^4 Z^2 
Z^{(4)}-\frac{1}{288} \l ^4 Z^4 Z^{(4)}+\frac{1}{720} \l ^4 Z^6 
Z^{(4)}+\frac{1}{120} \l ^5 Z Z' Z^{(4)}+\frac{1}{18} \l ^5 Z^3 Z' 
Z^{(4)}\lp
+\frac{1}{48} \l ^6 Z'^2 Z^{(4)}+\frac{11}{360} \l ^6 Z Z'' 
Z^{(4)}+\frac{\l ^5 Z^{(5)}}{30240}+\frac{1}{360} \l ^5 Z^2 
Z^{(5)}+\frac{1}{288} \l ^5 Z^4 Z^{(5)}\lp
+\frac{1}{72} \l ^6 Z Z' 
Z^{(5)}+\frac{19 \l ^6 Z^{(6)}}{51840}+\frac{1}{576} \l ^6 Z^2 
Z^{(6)}+\frac{\l ^7 Z^{(7)}}{10368}.
\ea

\section{Strong coupling asymptotic expansion of $\sF_{0}(\l)$}
\la{app:asymp}

From (\ref{5.13}), we see that the strong coupling expansion of $\sF_{0}(\l)$ amounts to the large $a\to \infty$ expansion of   
\be
f(a) = \frac{4}{\pi}\int_{0}^{1}dt\, \sqrt{1-t^{2}}\log\cosh(at) \qquad \to \qquad
f'(a) = \frac{2}{\pi}\int_{-1}^{1}dt\,t\, \sqrt{1-t^{2}}\tanh(at)
\ee
Expanding tanh, integrating, and replacing Bernoulli numbers by their integral representation gives
\be
f'(a) = -\frac{2\pi}{a^{2}}\int_{0}^{\infty}\frac{dt}{t}\frac{e^{2\pi t}}{(e^{2\pi t}-1)^{2}}[J_{1}(4at)-2J_{1}(2at)].
\ee
This has the form of a Mellin convolution
\be
f'(a) = -\frac{2\pi}{a}\int_{0}^{\infty}dt f_{1}(at)f_{2}(t) = -\frac{2\pi}{a}(f_{1}\star f_{2})(a),
\ee
with 
\be
f_{1}(t) = \frac{J_{1}(4t)-2J_{1}(2t)}{t}, \qquad f_{2}(t) = \frac{e^{2\pi t}}{(e^{2\pi t}-1)^{2}}
\ee
We have the explicit Mellin transforms
\ba
\mc M\left[\frac{e^{2\pi t}}{(e^{2\pi t}-1)^{2}}\right]  &= -\frac{1}{2\pi}\mc M\left[\frac{d}{dt}\frac{1}{e^{2\pi t}-1}\right] = \frac{1}{2\pi}(s-1) \mc M\left[\frac{1}{e^{2\pi t}-1}\right]_{s-1}\lp
=(2\pi)^{-s}\Gamma(s)\zeta(s-1),
\ea
and
\be
\mc M\left[ \frac{J_{1}(4t)-2J_{1}(2t)}{t}\right] = -\frac{2^{-s} \left(2^s-1\right) \Gamma \left(\frac{s}{2}\right)}{\Gamma
   \left(2-\frac{s}{2}\right)}.
\ee 
Hence, 
\ba
\mc M[f'](s) &= -2\pi \, \mc M[f_{1}\star f_{2}](s-1) = -2\pi\, \widetilde{f_{1}}(s-1)\widetilde{f_{2}}(2-s) \lp
= -\frac{2^{-s} \left(2^s-2\right) \zeta (s) \csc \left(\frac{\pi  s}{2}\right) \Gamma
   \left(\frac{s+1}{2}\right)}{\Gamma \left(\frac{5}{2}-\frac{s}{2}\right)},
\ea
and with $-1<c<0$
\be
f'(a) = -\int_{c-i\infty}^{c+\infty}\frac{ds}{2\pi i}a^{-s}\,\frac{2^{-s} \left(2^s-2\right) \zeta (s) \csc \left(\frac{\pi  s}{2}\right) \Gamma
   \left(\frac{s+1}{2}\right)}{\Gamma \left(\frac{5}{2}-\frac{s}{2}\right)}
\ee
Picking residues at poles with $s\ge  0$ we get the asymptotic expansion for $a\to \infty$. The first terms are
\be
f'(a) = \frac{4}{3\pi}-\frac{\pi}{6}\frac{1}{a^{2}}+\frac{7\pi^{3}}{480}\frac{1}{a^{4}}+\frac{31\pi^{5}}{16128}\frac{1}{a^{6}}+\frac{127\pi^{7}}{122880}\frac{1}{a^{8}}+\cdots,
\ee
with closed form 
\be
\la{E.10}
f'(a) =\frac{2}{\pi^{2}}\sum_{k=0}^{\infty}(1-2^{1-2k})\Gamma\left(k-\frac{3}{2}\right)\Gamma\left(k+\frac{1}{2}\right)\zeta(2k)\frac{1}{a^{2k}}.
\ee
This series is asymptotic and non-alternating so we expect non-perturbative corrections/ambiguities.
Integrating it we have 
\be
f(a) =C+\frac{2}{\pi^{2}}\sum_{k=0}^{\infty}\frac{1-2^{1-2k}}{1-2k}\Gamma\left(k-\frac{3}{2}\right)\Gamma\left(k+\frac{1}{2}\right)\zeta(2k)\frac{1}{a^{2k-1}}.
\ee
Since $\sF_{0}(\l)$ in (\ref{5.13}) obeys $\sF_{0}(\l) = f(\sql/2)$ we fix $C=-\log 2$.

\section{Solution of the differential equation  (\ref{5.19})}
\la{app:diff}

To solve the equation (\ref{5.19}), we set $f(x) = x h(\frac{1}{x^{2}})$ and get 
\be
1-e^{-2(h'+2x h'')}+4x h''=0.
\ee
We re-define $\log x = X$ and $H(X)=h'(x)$ and differentiate with respect to $X$
\be
(1+4H')' = -2(H'+2H'')e^{-2(H+2H')} = -2(H'+2H'')(1+4H').
\ee
Now set $H' = G$
\be
4G' = -2(G+2G')(1+4G).
\ee
The general solution is 
\be
G(X) = e^{-X/2}\,\bigg[c_{1}\pm\sqrt{c_{1}^{2}+\frac{1}{2}c_{1}e^{X/2}}\bigg].
\ee
Integrating back to get $H(X)\to h(x)\to f(x)$, we obtain 
\be
f(x) = -4c_{1}+c_{2}x+\frac{\sqrt{2c_{1}}}{3}\sqrt{2c_{1}+\frac{1}{x}}\,(5+4c_{1}x)+\frac{1}{x}\text{arccoth}\left(\frac{\sqrt{2c_{1}}}{\sqrt{2c_{1}+\frac{1}{x}}}\right)+\frac{i\pi}{2x}.
\ee
The integration constants are fixed by  the  expansion (\ref{5.18}) and we get the values
\be
c_{1}=\frac{1}{32\pi^{2}}, \qquad c_{3}=-\frac{1}{384\pi^{2}}.
\ee
After simplification, $f(x)$ takes the final form (\ref{5.20}).

\section{Systematic evaluation of $C_{n, m, \dots}^{SU(N)}$}
\la{app:Cn}

Let us begin with a single trace.
Splitting the trace in $U(N)$ we obtain 
\be
\vev{\tr e^{xA}}^{SU(N)} = e^{-\frac{x^{2}}{4N}}\,\vev{\tr e^{xA}}^{U(N)},
\ee
and taking derivatives
\be
\vev{A^{p}}^{SU(N)} = \frac{1}{2^{p}N^{p/2}}\vev{H_{p}(\sqrt N\, A)}^{U(N)}.
\ee
Using
\be
H_{2p}(x) = (2p)!\sum_{\ell=0}^{p}\frac{(-1)^{p-\ell}}{(2\ell)!(p-\ell)!}(2x)^{2\ell},
\ee
we obtain 
\be
C^{SU(N)}_{n} = (2n)!\sum_{\ell=0}^{n}\frac{(-1)^{n-\ell}}{(2\ell)!(n-\ell)!}\frac{1}{(4N^{2})^{n-\ell}}\,C_{\ell}^{U(N)}.
\ee
and we can read the systematic $1/N$ corrections, for instance
\be
\frac{C_{n}^{U(N)}-\frac{n(2n-1)}{2N^{2}}\,C_{n-1}^{U(N)}}{C_{n}^{SU(N)}} = 1+\mc O(1/N^{4}).
\ee
Multiple trace expectation values can be worked out similarly with the final result shown in (\ref{6.3}).

\section{A useful matrix identity}
\la{app:herm}

In (\ref{7.4}), we used the matrix identity
 \ba
d_{p}(A) = e^{\tr A^{2}}\partial_{A}^{p}e^{-\tr A^{2}} = (-1)^{p}N^{p/2} H_{p}(\tfrac{\tr A}{\sqrt N}).
 \ea
 This follows by direct inspection, or using the recursion 
 \ba
 d_{p+1}(A) &= e^{\tr A^{2}}\partial_{A}\partial_{A}^{p}e^{-\tr A^{2}} = e^{\tr A^{2}}\partial_{A}e^{-\tr A^{2}}e^{\tr A^{2}}\partial_{A}^{p}e^{-\tr A^{2}} \lp
 = (\partial_{A}-2\tr A)d_{p}(A).
 \ea
 Indeed, induction requires
 \be
-(-1)^{p}N^{p/2} \sqrt{N}H_{p+1}(\tfrac{\tr A}{\sqrt N}) = 
  (-1)^{p}N^{p/2}\bigg[\sqrt{N}H_{p}'(\tfrac{\tr A}{\sqrt N}) -2\tr A\, H_{p}(\tfrac{\tr A}{\sqrt N}) \bigg],
 \ee
and this holds using $H_{p}' = 2xH_{p}(x)-H_{p+1}(x)$.

\section{The large $N$ limit from the integral equation with sources}
\la{app:largeN}

In the large $N$ limit, the saddle point equation for the matrix model with generic defect function $L(x)$
 \be
 \la{I.1}
 \dashint_{-\ell}^{\ell}\,dv\,\frac{\rho(v)}{u-v} = 
 \frac{8\pi^{2}}{\l}\,u+\frac{1}{2N}\,L'(u),
 \ee
 where  $\ell=\frac{\sqrt\l}{2\pi}+\frac{1}{N}\delta \ell$ has a $\frac{1}{N}$
 correction to be determined. Let us rescale $u,v$ according to 
\be
u = \ell\,\alpha,\qquad \alpha\in[-1,1].
\ee
The integral equation (\ref{I.1}) reads \footnote{The new density is $\ell\,\rho(\ell\,\alpha)\to \rho(\alpha)$ such that $\int_{-1}^{1}d\alpha
\rho(\alpha)=1$.}
\be
\dashint_{-1}^{1}\,d\beta\,\frac{\rho(\beta)}{\alpha-\beta} = \frac{8\pi^{2}}{\l}\,\ell^{2}\,\alpha
+\frac{\ell}{2N}\,L'(\ell\,\alpha).
\ee
After expanding
\be
\rho(\alpha) = \frac{2}{\pi}\,\sqrt{1-\alpha^{2}}+\frac{1}{N}\,\delta\rho(\alpha)+\cdots,
\ee
the first order perturbed integral equation reads 
\be
\dashint_{-1}^{1}\,d\beta\,\frac{\delta\rho(\beta)}{\alpha-\beta} = \frac{8\pi}{\sqrt\l}\,\delta\ell\,\alpha
+\frac{\sqrt\l}{4\pi}\,L'\left(\frac{\alpha\sqrt\l}{2\pi}\right).
\ee
The bounded solution is unique and is obtained as 
\ba
\la{I.6}
& \delta\rho(\alpha) = -\frac{1}{\pi^{2}}\,\sqrt{1-\alpha^{2}}\,\dashint_{-1}^{1}d\beta\,\frac{1}{\alpha-\beta}
\frac{1}{\sqrt{1-\beta^{2}}}\,\bigg[
\frac{8\pi}{\sqrt\l}\,\delta\ell\,\beta
+\frac{\sqrt\l}{4\pi}\,L'\bigg(\frac{\beta\sqrt\l}{2\pi}\bigg)
\bigg], \notag \\
& \int_{-1}^{1}d\alpha\ \delta\rho(\alpha)=0.
\ea
where the second condition fixes $\delta\ell$.
The defect one-point function of $\mc O_{2n}$ is obtained from \footnote{Notice that for $2n=2$ we need to add to $T_{2}$ the constant shift $+\frac{1}{2}$, see \cite{Rodriguez-Gomez:2016cem}.}
\ba
& \VevD{  \tr T_{2n}\left(\frac{2\pi\,M}{\sqrt\l}\right) } = 
N\,\int_{-1}^{1}d\alpha\,\bigg[\frac{2}{\pi}\sqrt{1-\alpha^{2}}+\frac{1}{N}\delta\rho(\alpha)+\cdots\bigg]\,
T_{2n}\left(\frac{\frac{\sqrt\l}{2\pi}+\frac{\delta\ell}{N}+\cdots}{\frac{\sqrt\l}{2\pi}}\,\alpha\right) \lp
= N\,\int_{-1}^{1}d\alpha\,\bigg[\frac{2}{\pi}\sqrt{1-\alpha^{2}}+\frac{1}{N}\delta\rho(\alpha)+\cdots\bigg]\,\bigg[
T_{2n}(\alpha)+\frac{1}{N}\frac{2\pi\delta\ell}{\sqrt\l}T'_{2n}(\alpha)
\bigg], 
\ea
For $n>2$, using orthogonality of the Chebyshev polynomials, it is easy to show that the only surviving term is 
\be
\VevD{  \tr T_{2n}\left(\frac{2\pi\,M}{\sqrt\l}\right) } =
\int_{-1}^{1}d\alpha\,\delta\rho(\alpha)\,T_{2n}(\alpha).
\ee
Let us evaluate 
\ba
I_{n} &= \int_{-1}^{1}d\alpha\,\delta\rho(\alpha)\,T_{2n}(\alpha) = 
-\frac{\sqrt\l}{4\,\pi^{3}}\,\dashint_{-1}^{1}d\alpha\,\dashint_{-1}^{1}d\beta\,\frac{T_{2n}(\alpha)}{\alpha-\beta}
\frac{\sqrt{1-\alpha^{2}}}{\sqrt{1-\beta^{2}}}\,L'\left(\frac{\beta\sqrt\l}{2\pi}\right).
\ea
Let us consider the integral over $\alpha$
\ba
\dashint_{-1}^{1}d\alpha\,\,\frac{T_{2n}(\alpha)}{\alpha-\beta}
\sqrt{1-\alpha^{2}} = \dashint_{-1}^{1}d\alpha\,\frac{T_{2n}(\alpha)(1-\alpha^{2})}{\alpha-\beta}
\frac{1}{\sqrt{1-\alpha^{2}}} = \Asterisk
\ea
From the definition of Chebyshev polynomials, \cf Appendix \ref{app:cheb}, one obtains \footnote{This is also an immediate consequence of the product formula
$2\,T_{m}\,T_{n} = T_{m+n}+T_{|m-n|}$, 
evaluated at $m=2$.}
\be
-\frac{1}{4}T_{2n-2}(x)+\frac{1}{2}T_{2n}(x)-\frac{1}{4}T_{2n+2}(x) = T_{2n}(x)\,(1-x^{2}).
\ee
Hence, from (\ref{J.3}), we get 
\ba
\Asterisk &= -\frac{1}{4}\dashint_{-1}^{1}d\alpha\,\frac{T_{2n-2}-2T_{2n}+T_{2n+2}}{\alpha-\beta}
\frac{1}{\sqrt{1-\alpha^{2}}} = -\frac{\pi}{4}\bigg[U_{2n-3}(\beta)-2U_{2n-1}(\beta)+U_{2n+1}(\beta)\bigg] \notag \\
&= \pi\,(1-\beta^{2})\,U_{2n-1}(\beta).
\ea
Thus, 
\be
I_{n} = -\frac{\sqrt\l}{4\pi^{2}}\,\int_{-1}^{1}d\beta\,L'\left(\frac{\beta\sqrt\l}{2\pi}\right)\,\sqrt{1-\beta^{2}}\,U_{2n-1}(\beta),
\ee
and we obtain  (relabeling the integration variable)
\be
\la{I.14}
\OO_{n}^{(0)}(\l) = -\frac{2}{\pi}\,\bigg(\frac{\l}{16\pi^{2}}\bigg)^{\frac{n+1}{2}} \int_{-1}^{1}d\alpha\,L'\left(\frac{\alpha\sqrt\l}{2\pi}\right)\,
\sqrt{1-\alpha^{2}}\,U_{n-1}(\alpha).
\ee
Let us change variables $\alpha=\cos\theta$, using (\ref{J.1}) we get 
\ba
\OO^{(0)}_{n}(\l) &= -\frac{2}{\pi}\,\bigg(\frac{\l}{16\pi^{2}}\bigg)^{\frac{n+1}{2}} \int_{0}^{\pi}d\theta\,\sin^{2}\theta\,L'\left(\frac{\cos\theta\sqrt\l}{2\pi}\right)\,
\,U_{n-1}(\cos\theta) \lp
= -\frac{1}{\pi}\,\bigg(\frac{\l}{16\pi^{2}}\bigg)^{\frac{n+1}{2}} \int_{0}^{2\pi}d\theta\,\sin\theta\,L'\left(\frac{\cos\theta\sqrt\l}{2\pi}\right)\,
\,\sin(n\theta)  \lp
= \frac{1}{2\pi}\left(\frac{\l}{16\pi^{2}}\right)^{n/2}\int_{0}^{2\pi}d\theta\, \frac{d}{d\theta}L\left(\frac{\cos\theta\sqrt\l}{2\pi}\right)\,
\,\sin(n\theta) 
\ea
Integrating by parts
\ba
\OO^{(0)}_{n}(\l) &=  -\frac{n}{2\pi}\left(\frac{\l}{16\pi^{2}}\right)^{n/2}\int_{0}^{2\pi}d\theta\, L\left(\frac{\cos\theta\sqrt\l}{2\pi}\right)\,
\,\cos(n\theta)  \lp
= -\frac{n}{2\pi}\left(\frac{\l}{16\pi^{2}}\right)^{n/2}\oint \frac{dx}{ix}\, L\left(\frac{\sql}{4\pi}(x+x^{-1})\right)\,\frac{1}{2}(x^{n}+x^{-n}).
\ea
The terms $x^{n}$ and $x^{-n}$ give the same contribution using $x\to 1/x$ and we get (\ref{8.10}).

\section{Chebyshev polynomials}
\label{app:cheb}

Chebyshev polynomials of 1st and 2nd kind are defined by 
\begin{align}
\la{J.1}
& T_{n}(x) = \cos(n\theta),\quad U_{n}(x) = \frac{\sin[(n+1)\theta]}{\sin\theta}, \qquad 
\text{with}\ 
x= \cos\theta \in[-1,1],\ \ \theta\in[0,\pi],
\end{align}
and  obey the orthogonality relations
\begin{align}
\la{J.2}
& \int_{-1}^{1}dx \frac{T_{n}(x)T_{m}(x)}{\sqrt{1-x^{2}}} = \begin{cases}
0 & n\neq m \\ \pi & n=m=0 \\ \pi/2 & n=m \neq 0
\end{cases}, \qquad
\int_{-1}^{1}dx \sqrt{1-x^{2}}\,U_{n}(x)U_{m}(x) = \frac{\pi}{2}\delta_{nm}.
 \end{align}
Useful relations that we used in the main text are
\begin{align}
\la{J.3}
& T_{n}'(x) = n\,U_{n-1}(x),\notag \\  
& \int_{-1}^{1}dy\, \frac{T_{n}(y)}{(x-y)\,\sqrt{1-y^{2}}} = -\pi\,U_{n-1}(x),\qquad
 \int_{-1}^{1}dy\, \sqrt{1-y^{2}}\frac{U_{n}(y)}{x-y} = \pi\,T_{n+1}(x).
 \end{align}

\bibliography{BT-Biblio}
\bibliographystyle{JHEP-v2.9}
\end{document}